\documentclass[aps,pre,twocolumn,showpacs,superscriptaddress,floatfix]{revtex4-1}

\pdfoutput=1

\usepackage{latexsym}
\usepackage{amsmath,amsfonts,amssymb,amsbsy,amscd}
\usepackage{ifthen}
\usepackage[usenames,dvipsnames]{color}
\usepackage{hyperref}
\usepackage{graphicx}

\hypersetup{
   pdfauthor=Siminos,
   pdftitle=Effect of electron heating on relativistic self-induced transparency
}

\newcommand{\beq}{\begin{equation}}
\newcommand{\eeq}{\end{equation}}
\newcommand{\bseq}{\begin{subequations}}
\newcommand{\eseq}{\end{subequations}}

\newcommand{\rf}     [1] {~\cite{#1}}
\newcommand{\refref} [1] {Ref.~\cite{#1}}
\newcommand{\refrefs}[1] {Refs.~\cite{#1}}
\newcommand{\refeq}  [1] {Eq.~(\ref{#1})}
\newcommand{\refEq}  [1] {Equation~(\ref{#1})}
\newcommand{\refeqs} [2]{Eqs.~(\ref{#1})--(\ref{#2})}
\newcommand{\refEqs} [2]{Equations~(\ref{#1})--(\ref{#2})}
\newcommand{\refeqset} [2]{Eqs.~(\ref{#1})~and~(\ref{#2})}
\newcommand{\reffig} [1] {Fig.~\ref{#1}}

\newcommand{\refFig} [1] {Figure~\ref{#1}}

\newcommand{\refsect}[1] {Sec.~\ref{#1}}

\newcommand{\etal}{~{\em et al.}}
\newcommand{\ie}{{i.e.}}        
\newcommand{\cf}{{\em cf.\ }}   
\newcommand{\eg}{{e.g.\ }}

\newcommand{\ii}{i}

\renewcommand\Im{\operatorname{Im}}

\newcommand{\Epar}{\ensuremath{E_x}}
\newcommand{\ppar}{\ensuremath{p_x}} 
\newcommand{\pparo}{\ensuremath{\ppar{}_{0}}}
\newcommand{\dotppar}{\ensuremath{\dot{p}_x}} 
\newcommand{\pcr}{\ensuremath{\ppar^{\mathrm{cr}}}} 
\newcommand{\Hcr}{\ensuremath{H^{\mathrm{cr}}}} 
\newcommand{\pth}{\ensuremath{\ppar^{\mathrm{min}}}} %
\newcommand{\nceff}{\ensuremath{n_c^{\mathrm{eff}}}} 
\newcommand{\nC}{\ensuremath{n_{\mathrm{th}}}} 
\newcommand{\aC}{\ensuremath{a_{\mathrm{th}}}} 
\newcommand{\abC}{\ensuremath{a_{B}}} 
\newcommand{\tint}{\ensuremath{\tau_{\mathrm{int}}}} 
\newcommand{\psp}{\zeta} 
\newcommand{\IL}{\ensuremath{\mathrm{I}}} 
\newcommand{\tl}{\ensuremath{\tau_L}} 
\newcommand{\tr}{\ensuremath{\tau_r}} 

\newcommand{\eqb}[1]{\ensuremath{\mathrm{Q}_{#1}}} 
\newcommand{\jacElem}[3]{\ensuremath{A}_{#2#3}}
\newcommand{\jac}[1]{\ensuremath{\mathcal{A}}(\psp_#1)}

\newcommand{\rsit}{RSIT}
\newcommand{\csl}{CSL}
\newcommand{\cel}{CEL}
\newcommand{\squash}{\textsc{Squash}} 
\newcommand{\ri}{relativistically intense} 
\newcommand{\vf}{\ensuremath{v_f}}
\newcommand{\xEmax}{\ensuremath{x_b}}

\newcommand{\cweb}{(color online)}

\graphicspath{{figs/}}

\newcommand{\cwidth}{246.0pt} 

\definecolor{hreflinkcolor}{rgb}{0.13,0.17,0.83}
\hypersetup{colorlinks=true,urlcolor=hreflinkcolor,
		linkcolor=hreflinkcolor,citecolor=hreflinkcolor}

\begin{document}

\author{E. Siminos}
\email{evangelos.siminos@gmail.com}
\author{M. Grech}
\affiliation{Max Planck Institute for the Physics of Complex Systems, D-01187 Dresden, Germany}
\author{S. Skupin}
\affiliation{Max Planck Institute for the Physics of Complex Systems, D-01187 Dresden, Germany}
\affiliation{Friedrich Schiller University, Institute of Condensed Matter Theory and Optics, D-07743 Jena, Germany}
\author{T. Schlegel}
\affiliation{Helmholtz Institute Jena, D-07743 Jena, Germany}
\author{V. T. Tikhonchuk}
\affiliation{Univ. Bordeaux, CNRS, CEA, CELIA UMR 5107, F-33405 Talence, France}

\title{Effect of electron heating on self-induced transparency in relativistic intensity laser-plasma interaction}

\date{\today}

\begin{abstract}
The effective increase of the critical density associated with the interaction of relativistically intense laser pulses with overcritical plasmas, known as self-induced transparency, is revisited for the case of circular polarization.   A comparison of particle-in-cell simulations to the predictions of a relativistic cold-fluid model for the transparency threshold demonstrates that kinetic effects, such as electron heating, can lead to a substantial increase of the effective critical density compared to cold-fluid theory. These results are interpreted by a study of separatrices in the single-electron phase space corresponding to dynamics in the stationary fields predicted by the cold-fluid model. It is shown that perturbations due to electron heating exceeding a certain finite threshold can force electrons to escape into the vacuum, leading to laser pulse propagation. The modification of the transparency threshold is linked to the temporal pulse profile, through its effect on electron heating.   
\end{abstract}

\maketitle

\section{Introduction\label{s:intro}}

The optical properties of a plasma under the action of 
a relativistically intense laser pulse
(intensity $\IL\gtrsim10^{18}\,\mathrm{W\,cm^{-2}}$ for $1\,\mathrm{\mu m}$ wavelength)
are profoundly affected by nonlinearities in
the corresponding laser-plasma interaction.
In particular, the question of whether a pulse 
with the carrier frequency 
$\omega_L$ propagates in a 
plasma of electron density $n_0$ can no longer be 
answered solely in terms of the critical density,
\beq\label{eq:ncr} 
  n_c = \epsilon_0\,m_e\,\omega_L^2/e^2\,,
\eeq
where $m_e$ is the electron rest mass, $-e$ is the electron charge,  
and $\epsilon_0$ is the permittivity of free space. 
By definition, a \ri\ pulse accelerates electrons from rest 
to relativistic momenta within an optical cycle and, thus, 
the electron mass in \refeq{eq:ncr}
has to be corrected by the relativistic factor $\gamma=\sqrt{1+\mathbf{p}^2/m_e^2c^2}$,
where $\mathbf{p}$ is the electron momentum. For a purely transverse
wave propagating through a cold, homogeneous plasma,
this relativistic factor can be related, by the conservation of canonical momentum, 
to the normalized amplitude of the wave vector potential $a_0=eA_0/(m_ec)$, 
 $\gamma \simeq \sqrt{1+a_0^2/2}$~\footnote{The definition of the vector
potential is given in \refeq{eq:pulse}}. 
Therefore, one is forced to introduce an intensity-dependent effective 
critical density\rf{akhiezer1956,kaw1970}
\beq\label{eq:nceff}
  \nceff=\sqrt{1+\frac{a_0^2}{2}}\,n_c\,.
\eeq

According to \refeq{eq:nceff}, a \ri\ laser pulse ($a_0\gtrsim1$) 
can propagate through a nominally
overdense plasma, with electron density $n_c<n_e<\nceff$, 
a phenomenon known as \emph{relativistic self-induced transparency}
(\rsit). Apart from its role as a fundamental process in laser-plasma
interaction, \rsit\ is also interesting for applications, as it often determines the
regime of efficient laser-target interaction.
In the context of ion acceleration, for instance, \rsit\ can prevent
efficient ion radiation-pressure-acceleration from thin targets\rf{klimo_PRSTAB_2008,Robinson_NJP_2008,Yan_PRL_2008,
Macchi_PRL_2009,Grech_NJP_2011}
or laser-driven hole-boring in thicker ones\rf{Naumova_PRL_2009,Schlegel_POP_2009}. 
On the other hand, \rsit\ may enhance electron heating in the break-out afterburner 
acceleration 
mechanism, thus allowing for higher ion energies\rf{Yin_LPB_2006,Albright_POP_2007,Yin_PRL_2011}.

In this paper, we investigate \rsit\ in the 
case of a circularly polarized (CP) laser pulse with finite rise (or ramp-up) time $\tau_r$
and infinite duration, normally incident onto a semi-infinite 
plasma with a constant density $n_0 > n_c$, and a sharp interface
with the vacuum. This configuration is of particular interest for ultrahigh 
contrast
laser interaction with thick targets.
Unfortunately, the simple relation \eqref{eq:nceff}, derived 
assuming a purely transverse 
plane-wave and a homogeneous plasma of infinite extent, does not apply
to this setting. The main reason for this is that the effect of the ponderomotive force 
(associated here with inhomogeneities along the propagation direction) 
becomes dominant and leads to a significant modification of \rsit\ threshold.
Since the 1970's, several analytical studies, mostly within the framework 
of relativistic, cold-fluid theory\rf{akhiezer1956}, have been undertaken to investigate
strong electromagnetic wave propagation through inhomogeneous 
plasmas\rf{max1971, marburger1975, Lai1976},
culminating in a derivation of a modified \rsit\ threshold which
incorporates boundary conditions at the plasma-vacuum 
interface\rf{cattani2000, goloviznin2000}. 
In order to establish contact with this line of previous work and to focus on the key physical
mechanisms, we will restrict attention to immobile ions 
and one--dimensional geometry.
%
\begin{figure*}
 \includegraphics{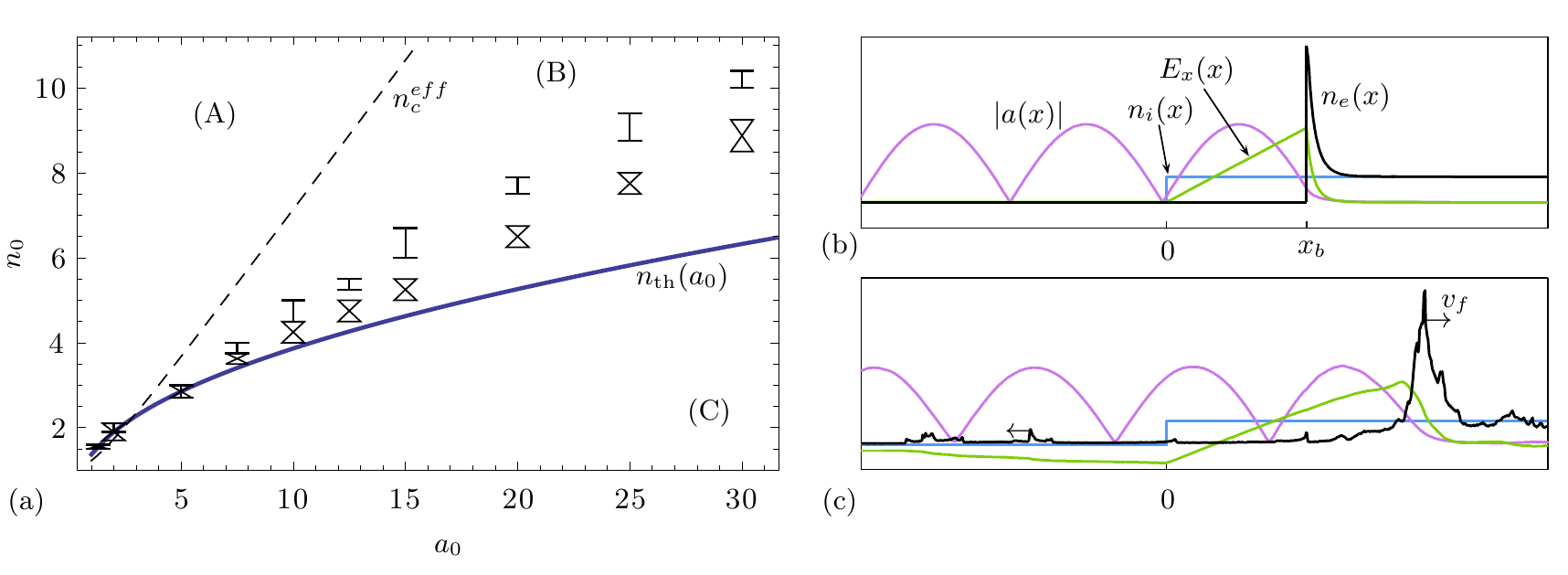}
  \caption{\cweb\
	(a)~Effective critical density as a function of the laser
	field amplitude $a_0$ as predicted by
	the simple relation \eqref{eq:nceff} (dashed, black line). 
	Threshold density $\nC(a_0)$ below
	which, according to the cold-fluid theory (\cf \refsect{s:review}),  
	no standing wave solutions exist (solid, blue line).
	\rsit\ threshold as extracted from our PIC simulations 
	(\cf \refsect{s:numerics}) with two different pulse rise times: 
	$0.25\tau_L$ (error bars) and $4\tau_L$ (triangular error bars), 
	where $\tau_L=2\pi/\omega_L$ is the laser period.
	(b)~Schematic representation of the stationary solution
	predicted by the cold-fluid theory for the case of total reflection 
	[regions (A) and (B) of panel (a)]. 
 	Shown are the electric field $\Epar(x)$, 
 	vector potential of the standing wave $|a(x)|$, 
 	and ion (electron) density $n_i(x)$ [$n_e(x)$], see \refsect{s:review}
 	for details. 
	(c)~Schematic representation of a typical case of pulse 
 	propagation in PIC simulations 
	[for \rsit\ in region (B) or (C) of panel (a)].
 	Arrows indicate the direction of electron motion.
 	See \refsect{s:numerics} for numerical results.
  }
  \label{f:threshold_geom}
\end{figure*}

Based on the assumptions stated above, 
the relativistic cold-fluid model predicts total reflection of 
the incident pulse, if a certain density threshold $\nC(a_0)$ 
is exceeded\rf{cattani2000,goloviznin2000} [see solid blue line 
in \reffig{f:threshold_geom}(a)].
The geometry of the stationary state predicted for $n_0>\nC(a_0)$ 
is illustrated in \reffig{f:threshold_geom}(b):
the ponderomotive force pushes the electrons deeper into the plasma, creating a charge 
separation layer (\csl) and an electron density spike [henceforth referred to as
compressed electron layer (\cel)] at the edge of the plasma.
Electrons in the \cel\ experience a strong electrostatic field (due
to charge separation), which balances the ponderomotive force.
The density in the \cel\ is typically much higher than 
$\nceff$ and, thus, pulse propagation is inhibited
and a standing wave is formed.

For plasma densities $n_0<\nC(a_0)$, such stationary solutions cease 
to exist, and one enters the regime of \rsit. 
particle-in-cell (PIC) simulations\rf{goloviznin2000,eremin2010}, however, indicate that 
light propagation in this regime is quite 
different from the traveling-wave solutions discussed earlier\rf{kaw1970}.
Although a \cel\ is initially formed, electrons at its edge escape 
toward the vacuum, leading to force imbalance and allowing the ponderomotive
force to push the \cel\ deeper into the target.
The situation is more reminiscent of hole-boring~\rf{wilks1993,Schlegel_POP_2009} 
(albeit with immobile ions) with a penetration front
moving deeper into the plasma with a constant velocity $v_f$, 
and a Doppler-shifted reflected wave [\reffig{f:threshold_geom}(c)] (see also 
\refrefs{lefebvre1995, guerin1996,sakagami1996, gibbon2005}).

In this work we show, through PIC simulations, that in the 
presence of electron heating, 
induced by the pulse finite rise time,
such a propagation mechanism can be activated even for densities
$n_0>\nC(a_0)$; see \reffig{f:threshold_geom}(a). 
The crucial role is again played by electrons at the
edge of the \cel\ escaping toward the vacuum. 
However, it has been recently shown that, 
in the total reflection regime, 
electrons at the edge of the \cel\ cannot be
forced to escape into the vacuum by \emph{infinitesimal} perturbations\rf{eremin2010}.
To interpret our results, we are thus led to study the response
of electrons at the edge of the \cel\ to \emph{finite}
perturbations. 
Studying the dynamics of a test-electron in the stationary fields
predicted by the cold-fluid model for the \csl\ and vacuum,
we show that electron escape to the vacuum is controlled
by separatrices in the single-electron phase space.
Moreover, we demonstrate that the perturbation threshold 
for unbounded motion (electron escape) predicted by our analytical considerations
is comparable to the attainable electron momentum 
due to heating (in the \cel), observed in our PIC simulations at the threshold for RSIT.
Finally, we study the effect of laser pulse rise time 
on electron heating and on the observed modification of the \rsit\ threshold.

This paper is organized as follows. In \refsect{s:review} we revisit some results
of the stationary cold-fluid theory that motivate
the present study.
In \refsect{s:PS}
we analyze the single-electron phase space (for motion in vacuum and charge
separation layer), by determining equilibrium solutions (\refsect{s:eqb}), studying
their linear stability (\refsect{s:stab}), and determining separatrices of bounded and
unbounded motion (\refsect{s:sep}). In \refsect{s:numerics} we present our PIC 
simulation results and relate them to the analytical results of \refsect{s:PS}.
Finally, we discuss our findings and present our conclusions in \refsect{s:disc}.

\section{Review of relativistic cold-fluid theory for \rsit\label{s:review}}

Throughout the paper, all quantities are normalized to 
(so-called) relativistic units. In particular, velocity, time, and distance are 
normalized to the speed of light $c$, inverse laser frequency $\omega_L^{-1}$, 
and inverse vacuum wave number $k_L^{-1} = c/\omega_L$, respectively.  
Electric charges and masses are normalized to $e$ and $m_e$, respectively, 
and densities are 
normalized to the critical density $n_c$. 
Finally, electric fields are normalized to the Compton field $E_C = m_e\,c\,\omega_L/e$.

\subsection{Stationary cold plasma model\label{s:cattani}}

In this section, we revisit the one-dimensional stationary model proposed 
independently by 
Cattani\etal\rf{cattani2000} and Goloviznin and Schep\rf{goloviznin2000} 
to describe the reflection of an incident
relativistic CP laser pulse by a nominally overdense 
plasma with constant electron density $n_0>1$ and a sharp interface
with vacuum. Our presentation follows
\refref{cattani2000}.

We consider an incident CP laser pulse propagating along the ${\bf \hat{x}}$-direction
with the vector potential
\beq\label{eq:pulse}
{\bf A}_L(t,x) = \frac{a_0}{\sqrt{2}}\,\left[\cos(t-x)\,{\bf \hat{y}} 
         + \sin(t-x)\,{\bf \hat{z}} \right]\,,
\eeq
where ${\bf \hat{y}}$ and ${\bf \hat{z}}$ denote the unit vectors 
forming an orthonormal basis in the plane transverse to the laser 
propagation direction.
The pulse is incident from vacuum ($x<0$) onto a semi-infinite plasma ($x>0$).
In this work, as in \refrefs{goloviznin2000,cattani2000}, 
we will neglect ion motion. 

As outlined in the Introduction, we will consider stationary solutions expressing
the balance of the ponderomotive and electrostatic forces, achieved once a \csl\ 
of sufficient thickness $x_b$ is created, see \reffig{f:threshold_geom}(b). 
Assuming total reflection of the laser pulse by the plasma, 
the balance of the  radiation ($\sim a_0^2$) and electrostatic
pressures [$\sim(n_0\,x_b)^2/2$], provides a rough estimate for 
the thickness of the \csl,
\beq\label{eq:xb_approx}
  x_b \simeq \frac{\sqrt{2}\,a_0}{n_0}\,.
\eeq
The exact expression for $x_b$ and the limits of applicability of \refeq{eq:xb_approx} are discussed below; see \refeq{eq:xb}.

In the following, we will look for stationary solutions with vector potential of the form
\beq\label{eq:standing}
{\bf A}(t,x) = a(x)\left[\cos\left(t+\theta/2\right)\,{\bf \hat{y}}+\sin\left(t+\theta/2\right)\,{\bf \hat{z}}\right],
\eeq
where $\theta$ accounts for the phase jump of the reflected wave 
\[
{\bf A}_R(t,x)=\frac{a_0}{\sqrt{2}}\,\left[\cos(t+x+\theta)\,{\bf \hat{y}} 
         + \sin(t+x+\theta)\,{\bf \hat{z}} \right]
\]
at $x=x_b$ and will be computed below.
In what follows, we will refer to the spatial function $a(x)$
as the ``vector potential.'' Note that, in the absence of plasma, 
we have $a(x)=\sqrt{2}\, a_0 \cos\left(x+\theta/2\right)$.

Modeling electrons as a relativistic cold fluid, as in \refref{cattani2000},
we seek stationary solutions satisfying the system of equations
\begin{align}
  \label{eq:m1} \frac{d\phi}{dx}    &= \frac{d\gamma}{dx}\,, \\
  \label{eq:m2} \frac{d^2\phi}{dx^2}&= n_e-n_0\,,\\
  \label{eq:m3} \frac{d^2a}{dx^2}   &= \left(\frac{n_e}{\gamma}-1\right)\,a\,.
\end{align}
Here, $\phi(x)$ is the electrostatic potential, $n_e(x)$ is the electron 
density and the Lorentz factor is written as
$\gamma(x) = \sqrt{1+a^2(x)}$ through conservation of transverse
canonical momentum. 
\refEq{eq:m1} expresses the balance between the electrostatic and 
ponderomotive forces inside the plasma. Hence, it holds only 
for $x \ge x_b$. 
\refEq{eq:m2} is simply Poisson equation and \refeq{eq:m3} 
is the propagation equation (in the Coulomb gauge) for the
field prescribed by \refeq{eq:standing}.

To solve the system of \refeqs{eq:m1}{eq:m3}, one considers
the \csl\ and the \cel\ separately.
The electron density $n_e(x)$, electrostatic field $\Epar(x)=-d\phi/dx$ and vector 
potential $a(x)$ are obtained in each layer. Solutions are then matched at 
the electron front $x=x_b$ to ensure continuity of $a(x)$, 
of its first derivative $da/dx$ and of $\Epar(x)$.

\subsection{Charge separation layer\texorpdfstring{, $0 \le x \le x_b$}{}}

The electrostatic field in the \csl, $0 \le x \le x_b$, 
is easily found by integrating Poisson \refeq{eq:m2} with
$n_e=0$ (no electrons) and boundary condition
$\Epar(0)=0$ (to match the electrostatic field at the
vacuum),  
\beq\label{eq:ex_csl}
  \Epar(x) = -\frac{d\phi}{dx} = n_0\,x\,.
\eeq
Thus, the electrostatic field for $0 \le x \le x_b$ increases
linearly, up to its maximum value $E_b \equiv \Epar(x_b) = n_0\,x_b$.
For total reflection at $x=x_b$ we can integrate \refeq{eq:m3} once to get
\beq\label{eq:TR}
  \left. \left(\frac{da}{dx}\right)^2\right\vert_{x=x_b} = 2\,a_0^2 - a_b^2\,,
\eeq
where $a_b=a(x_b)$ is the vector potential at the plasma boundary. Here, we write
the amplitude of the standing wave arising from the combination of the incident and reflected waves $A_L$ and $A_R$, respectively, as
\beq\label{eq:a_csl}
  a(x) = \sqrt{2}\,a_0\,\sin\left[\arcsin\left(\frac{a_b}{\sqrt{2}\,a_0}\right) - (x-x_b) \right]\,,
\eeq
which implies that in \refeq{eq:standing} we have $\theta/2=\pi/2-\arcsin(a_b/\sqrt{2}\,a_0)-x_b$.
At this point, there are two unknown quantities, $x_b$ and $a_b$, 
which will be determined self-consistently by considering the region $x \ge x_b$.
Note that we assume $a_b>0$, while from \refeq{eq:a_csl} we have $a'(x_b)<0$ so that the 
ponderomotive force $d\gamma/d\,x=\gamma^{-1}\,a\,da/dx$ pushes electrons deeper into the plasma, thus 
balancing the electrostatic force.

\subsection{Compressed electron layer\texorpdfstring{, $ x \ge x_b$}{}}

We now derive equations for the electron density, vector potential 
and electrostatic field in the plasma, $x \ge x_b$.
Combining \refeqset{eq:m1}{eq:m2}, one can rewrite the normalized electron density 
in the plasma as a function of the vector potential $a(x)$ and its first 
two derivatives,
\beq\label{eq:n_cel}
  n_e(x) = n_0 + \frac{1}{\sqrt{1+a^2}}\,\left[ a\,\frac{d^2 a}{dx^2} + \frac{1}{1+a^2}\,\left(\frac{da}{dx}\right)^2 \right].
\eeq
Substituting \refeq{eq:n_cel} in \refeq{eq:m3}, we obtain a differential 
equation for the vector potential only:
\beq\label{eq:diff_cel}
  \frac{d^2 a}{dx^2}=\frac{a}{1+a^2}\!\left(\frac{da}{dx}\right)^2\!-\!\left(1\! +\!a^2 \!- \!n_0\sqrt{1+a^2}\right)\!a\,.
\eeq

In the case of total reflection, \refeq{eq:diff_cel} describes 
the evanescent field in the 
overdense plasma, and has to be solved with boundary conditions
$a(x)\rightarrow 0$ and $da/dx\rightarrow 0$ 
for $x \rightarrow + \infty$~\footnote{\refEq{eq:n_cel} then implies
$n_e(x)\rightarrow n_0$, 
as $x\rightarrow\infty$.}.
\refEq{eq:diff_cel} admits a first integral,
\beq\label{eq:a_cel_int}
  \frac{1}{2(1+a^2)}\left(\frac{d\,a}{d\,x}\right)^2-\frac{1}{2}\left(2n_0\sqrt{1+a^2}-a^2\right)=-n_0,
\eeq 
which may be used to derive a 
solution that satisfies 
the required boundary conditions~\cite{marburger1975},
\beq\label{eq:a_cel}
  a(x) = \frac{2 \sqrt{n_0\,(n_0-1)}\,\cosh\left[(x-x_0)/\lambda_s\right]}{n_0\,\cosh^2\left[(x-x_0)/\lambda_s\right]-(n_0-1)}\,,
\eeq
where $\lambda_s = (n_0-1)^{-1/2}$ is the classical skin-depth, and $x_0$ 
is determined by ensuring the continuity of the vector potential at $x = x_b$. 

With $a(x)$ inside the plasma provided by \refeq{eq:a_cel}, 
one obtains $n_e(x)$ from \refeq{eq:n_cel}, while
\refeq{eq:m1} provides the electrostatic field in this region,
\beq\label{eq:ex_cel}
  \Epar(x) = -\frac{d}{dx}\sqrt{1+a^2}\,.
\eeq
\refEq{eq:ex_cel} together with \refeqset{eq:ex_csl}{eq:TR} and the
continuity of the electrostatic field at $x=x_b$ gives 
an explicit expression for the position of the electron front,
\begin{eqnarray}\label{eq:xb}
x_b = \frac{a_b}{n_0}\,\sqrt{\frac{2\,a_0^2 - a_b^2}{1+a_b^2}}\,.
\end{eqnarray}

Finally, from \refeqset{eq:TR}{eq:a_cel_int}, one obtains:
\beq\label{eq:ab}
  \frac{2\,a_0^2 + a_b^4}{1+a_b^2} = 2\,n_0\,\left(\sqrt{1+a_b^2} - 1 \right)\,.
\eeq
This equation defines, for a given incident laser field amplitude 
$a_0$ and initial plasma density $n_0$, the maximum evanescent field $a_b$ 
in the plasma. Solutions $a_b$ of \refeq{eq:ab} should satisfy 
the additional condition
\beq\label{eq:cond_ab}
  2\,a_0^2 - a_b^2 \geq 0\,,
\eeq
which follows from \refeq{eq:TR}.

Note that, in the limit $1 \ll a_b \ll a_0$, \refeq{eq:xb} allows us 
to recover the approximate result \refeq{eq:xb_approx}. 
On the other hand, from \refeqs{eq:xb}{eq:ab} we find that 
in the limit $a_b \ll 1$ (correspondingly $a_0^2 \ll n_0$) 
$x_b \simeq 2\,a_0^2/n_0^{3/2}$.

\subsection{Threshold for \rsit}

For a given plasma density $n_0$, \refeq{eq:ab} admits a solution only when 
the maximum evanescent field $a_b$ satisfies\rf{cattani2000}
\beq\label{eq:stab_0}
  2\,(n_0 + a_b^2) \le 3\,n_0\,\sqrt{1+a_b^2}\,.
\eeq
As shown in \refref{goloviznin2000}, 
for $n_0<3/2$, solutions compatible with \refeq{eq:cond_ab} 
can only be found in the region $a_0^2\leq2\,n_0(n_0-1)$. 
Thus, in this case, the threshold
incident laser amplitude reads
\beq\label{eq:a_threshold2}
  \aC^2 = 2\,n_0(n_0-1)\,.
\eeq

For $n_0>3/2$ condition \eqref{eq:cond_ab}
is always fulfilled and \refeq{eq:stab_0} defines the regime of total reflection. 
The threshold for \rsit\ corresponds to equality in \refeq{eq:stab_0}.
The maximum evanescent field at the threshold then reads
\beq\label{eq:ab_threshold}
  \abC^2 = n_0\,\left( \frac{9}{8}\, n_0 - 1 + \frac{3}{2}\, \sqrt{\frac{9}{16}\, n_0^2 - n_0 + 1} \right)\,.
\eeq
The threshold incident laser field amplitude $a_{th}$ above which \rsit\ occurs
in a plasma with initial density $n_0$ is obtained by substituting $a_b=\abC$
from \refeq{eq:ab_threshold} in \refeq{eq:ab},
\beq\label{eq:a_threshold}
  \aC^2 = n_0\,(1+\abC^2)\,\left(\sqrt{1+\abC^2} -1 \right) - \abC^4/2\,.
\eeq

Depending on the density range, \refeq{eq:a_threshold2}, respectively \refeqs{eq:ab_threshold}{eq:a_threshold},
define a threshold amplitude $\aC(n_0)$, above which \rsit\ occurs, 
for a given plasma density. Alternatively, for a 
given incident amplitude $a_0$, we may read \refeq{eq:a_threshold2}, respectively
\refeqs{eq:ab_threshold}{eq:a_threshold}, 
as defining an effective critical density 
$\nC(a_0)$ below which \rsit\ occurs. 
This is illustrated in \reffig{f:threshold_geom}(a).
\refEq{eq:a_threshold2} yields
\beq
  \nC(a_0)= \frac{1}{2}\left(1 + \sqrt{1 + 2 a_0^2}\right)\,,\quad \nC<3/2\,,
\eeq 
while 
\refeqs{eq:ab_threshold}{eq:a_threshold} can be inverted analytically 
in the limit $n_0\gg1$, 
yielding
\beq\label{eq:nCappr}
  \nC(a_0) \simeq\frac{2}{9} \left(3 + \sqrt{9 \sqrt{6}\, a_0 - 12}\,\right)\,,\quad \nC\gg1\,.
\eeq
Thus, the asymptotic behavior of $\nC(a_0)$ in the limit $a_0\gg1$ is $\nC\propto a_0^{1/2}$,
a much more restricting condition than \refeq{eq:nceff}, which for large $a_0$ becomes
$\nceff\propto a_0$.

As discussed in the Introduction, our PIC simulations indicate that
for pulses with finite rise time, the transition between total reflection 
and \rsit\ occurs within the limits set by \refeqset{eq:nCappr}{eq:nceff} and, 
moreover, depends on the pulse rise time.
In order to explain this discrepancy, we will now study
single electron dynamics in the stationary fields (in vacuum and \csl)
calculated above.

\section{Single electron dynamics\label{s:PS}}

\subsection{Equations of single electron motion\label{s:hamiltonian}}

The equations of motion for an electron in the region $x \le x_b$ (\ie\ in
the vacuum and \csl), in the case
of total reflection, read
\begin{align}
  \dot{x} & = \ppar/\gamma\,,\label{eq:dotx}\\
  \dotppar & = -\frac{\partial\,\gamma}{\partial\,x}  -  \Epar(x)\,,\label{eq:dotp}
\end{align}
where we have used conservation of transverse canonical momentum to write the electron $\gamma$ factor as
\beq
  \gamma(x,\ppar)=\sqrt{1+a^2(x)+\ppar^2},
\eeq
$\ppar$ is the electron's longitudinal momentum, the electrostatic
field $\Epar(x)$ and vector potential $a(x)$ are given by 
\refeqset{eq:ex_csl}{eq:a_csl}, respectively, and dotted quantities
are differentiated with respect to time.  

\refEqs{eq:dotx}{eq:dotp} can be derived from the Hamiltonian:
\beq\label{eq:H}
  H(x,\ppar)=\gamma(x,\ppar)-\phi(x)\,,
\eeq
where the electrostatic potential reads
\beq\label{eq:phi}
  \phi(x)=\begin{cases}
    0\,, &  x<0\,,\\
    -\frac{1}{2}n_0\, x^2, &  0 \le x \le x_b\,.
           \end{cases}
\eeq

\begin{figure}[htpb]
  \begin{center}
    \includegraphics[width=\cwidth, clip=true]{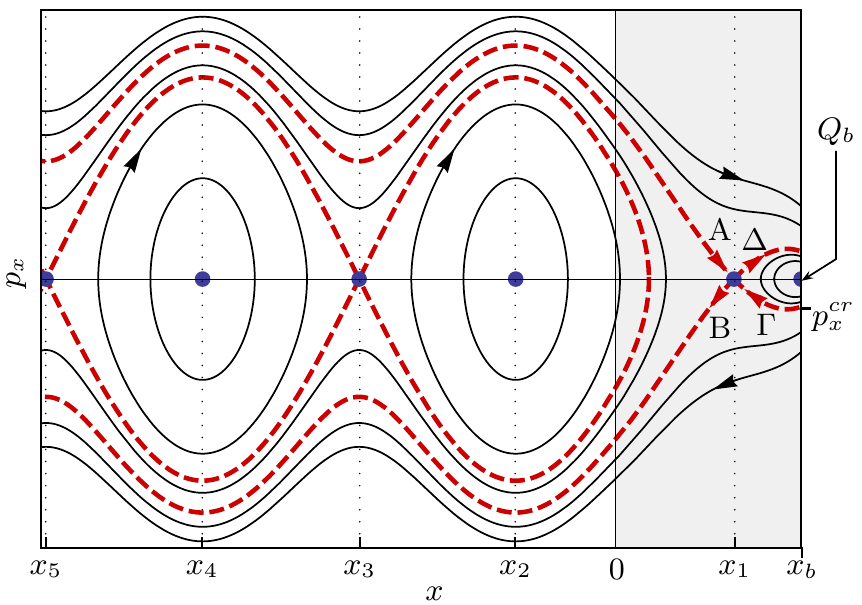}
  \end{center}
  \caption{\cweb\ Typical single-electron phase space portrait for \refeqs{eq:dotx}{eq:dotp}. 
	The first six equilibria $\eqb{b},\, \eqb{1},\,\ldots,\,\eqb{5}$ 
	corresponding to positions $x_b,\,x_1,\,\ldots,\,x_5$ and zero momentum
	are shown as blue dots. Separatrices are shown as red, 
	dashed lines and some typical
	trajectories are depicted as black, solid lines. The CSL is depicted
	as a gray-shaded area.
  }
  \label{f:PS}
\end{figure}

The Hamiltonian $H(x,\ppar)$ is a conserved quantity and we can thus write 
an explicit expression for the electron orbit with
initial conditions $x_0,\pparo$:
\beq \label{eq:PSorb}
  \ppar(x)=\pm\,\sqrt{\left[H(x_0,\pparo)+\phi(x)\right]^2-a^2(x)-1}\,.
\eeq
\refEq{eq:PSorb} suffices to plot portraits of the single-electron phase space,
as shown in \reffig{f:PS}. In the following subsections we explain how 
the several solutions
depicted in \reffig{f:PS} are interrelated, in order to understand 
how phase space geometry 
affects the threshold of \rsit. 
%
%

\subsection{Equilibrium solutions\label{s:eqb}}

The simplest type of solutions of \refeqs{eq:dotx}{eq:dotp} are equilibrium 
solutions for which $\dot{x}=\dotppar=0$.
We have already seen that, within the framework of the stationary cold-fluid model, 
the force balance \refeq{eq:m1} is satisfied in the plasma 
and in particular at 
$x=x_b$. Thus, the point $(x,\ppar)=(x_b,0)$ is an equilibrium which
we label as $\eqb{b}$. (For the same reasons, any point in the plasma 
with $\ppar=0$ will be an equilibrium.)

In the \csl\ and vacuum, on the other hand, the ponderomotive and electrostatic
forces are not balanced in general,
and equilibria for the motion of a test particle have to be found by setting 
$\dot{x}=\dotppar=0$ in \refeqs{eq:dotx}{eq:dotp}. We label equilibria at the 
left of $\eqb{b}$ as $\eqb{m}$, $m=1,\,2,\dots$, 
where $m$ increases with decreasing $x_m$.

For $x<0$ (in the vacuum) equilibria correspond to
$\partial_x \gamma=\frac{a}{\gamma}\frac{da}{dx} = 0$, \ie\ $a=0$ or $da/dx=0$, 
which, according to \refeq{eq:a_csl}, leads to
\beq\label{eq:eqbVac}
  x_k^- = \arcsin\left(\frac{a_b}{\sqrt{2}\,a_0}\right) + x_b - k \pi/2\,.
\eeq
Here, $k$ can be any positive integer provided that $x_k^-<0$, and $k$ even or odd 
correspond to $a(x_k^-)=0$ or $a'(x_k^-)=0$, respectively. 
We note that in our
labeling scheme, index $k$ in $x_k^-$ does not always correspond to index $m$
in labeling of equilibria \eqb{m}, \ie\ we will generally have $x_m=x_k^-$ with
$m\neq k$.~\footnote{The difference $m-k$ corresponds to the number 
of equilibria in the \csl, which is \emph{a priori} unknown.}

\begin{figure*}
  \begin{center}
	\includegraphics[clip=true]{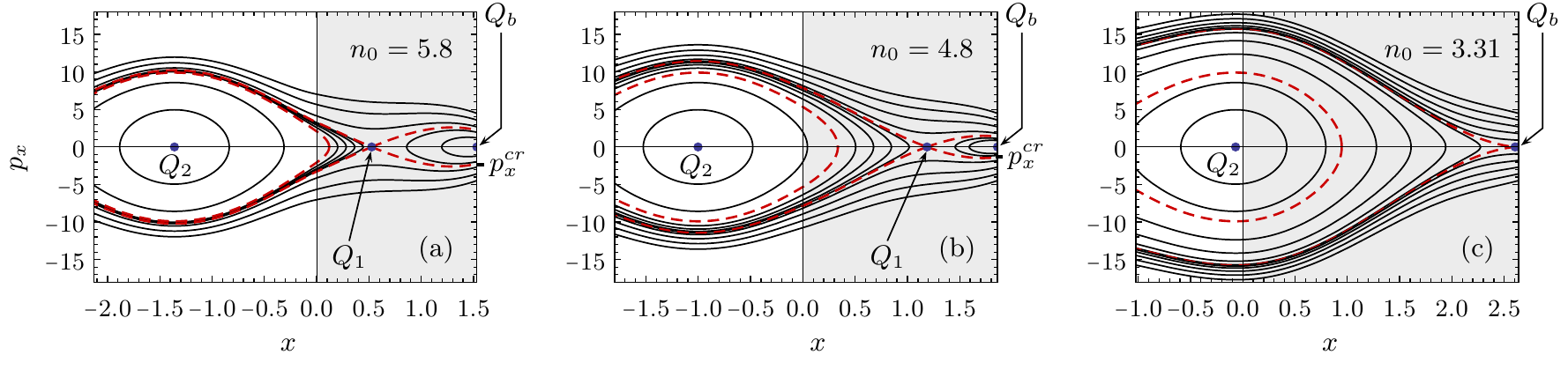}
  \end{center}
  \caption{\cweb\ For a given laser field amplitude, here $a_0=7$, 
	the absolute value of the 
	critical momentum for an electron to escape to the vacuum
	$|\pcr|$ decreases as $n_0$ decreases. 
	Shown are the cases (a) $n_0 = 5.8$, (b) $n_0 = 4.8$, (c) $n_0 = 3.31$.
	Color-code is the same as described in the caption of \reffig{f:PS}. 
	In panel (c) equilibria $\eqb{1}$ and $\eqb{b}$ cannot be distinguished
	within the resolution of this plot, $n_0$ being 
	slightly above the cold-fluid theory threshold $\nC=3.30458$.
	Note that the scale of $x$ and $\ppar$ has been kept the same in all panels.
  }
  \label{f:sep}
\end{figure*}

For $0\leq x \leq x_b$ (in the \csl), the equilibrium condition 
$\partial_x\phi=\partial_x\gamma$ must be solved numerically,
using \refeqset{eq:a_csl}{eq:phi} for $a(x)$ and $\phi(x)$,
respectively.
A perturbative solution can be obtained in the neighborhood of $x_b$, 
by expanding $\dot{p}=-\frac{\partial\,\gamma}{\partial\,x}  -  \Epar(x)=0$ 
to second order in $x-x_b$.
We obtain two solutions, $x=x_b$ and
\beq\label{eq:x1}
  x_1 \simeq x_b +\frac{2(1+a_b^2)^2 [2(a_b^2+n_0)-3\,n_0(1+a_b^2)^{1/2}]}{a_b(2a_0^2-a_b^2)^{1/2}(4+2a_b^2+a_b^4+6a_0^2)}\,.
\eeq
Comparing \refeq{eq:x1} with condition \eqref{eq:stab_0}, we see that 
$x_1\leq x_b$, as long as a standing wave solution exists, 
\ie\ for $n_0\geq\nC$. 
At threshold, $n_0=\nC$, we have $x_1=x_b$. 
That is, if we approach the \rsit\ threshold (as predicted by cold-fluid theory), 
the equilibrium \eqb{1} approaches \eqb{b} until the two states coalesce,
see \reffig{f:sep}(c).
%

\subsection{Stability\label{s:stab} of equilibria}

Linear stability analysis of the equilibria determined in \refsect{s:eqb} 
can give us
information on the behavior of orbits in the neighborhood of the
equilibria. 
For notational convenience, we define phase space variables 
$\psp=(\psp_1,\psp_2)\equiv(x,\, \ppar)$ and
rewrite the equations of motion [\refeq{eq:dotx} and \refeq{eq:dotp}] in the form
\beq\label{eq:eomFormal}
 \dot{\psp}_i = F_i(\psp)\,.
\eeq
where  $F_1(\psp) = \psp_2/\gamma$ and 
$F_2(\psp)=-\frac{\partial\,\gamma}{\partial\,\psp_1}  -  \Epar(\psp_1)$.
Considering infinitesimal perturbations in the neighborhood of equilibrium $\psp^{(m)}$,
and substituting $\psp(t)=\psp^{(m)}+\xi(t)$, with $\lVert\xi\rVert\ll1$, in \refeq{eq:eomFormal},
one obtains
\beq\label{eq:linear}
  \dot{\xi} = \jac{m}\xi\,,
\eeq
where the Jacobian matrix $\jac{m}$, with elements
\beq\label{eq:jacDef}
 \jacElem{m}{i}{j}=\left.\frac{\partial F_i}{\partial \psp_j}\right|_{\eqb{m}}\,,
\eeq
has been introduced.

Solutions of the linear system \refeq{eq:linear} are of the form 
$\xi(t)=\exp[\jac{m}t]\xi(0)$, and thus the linear stability 
of equilibrium $\eqb{m}$ is determined 
by the eigenvalues of the Jacobian matrix.
In Hamiltonian systems with one degree of freedom, classification of 
equilibria \eqb{m} by linear stability is straightforward 
(see, e.g., \refref{lichtenberg1992}), as there are only two possibilities:
\begin{itemize}
 \item $\jac{m}$ has a pair of real eigenvalues $\lambda_1=-\lambda_2>0$.
		Solutions then deviate from \eqb{m} at an exponential rate, 
		$\lVert\xi(t)\rVert\sim e^{\lambda_1 t}\lVert\xi(0)\rVert$, 
		and the equilibrium (called a \emph{saddle}) is unstable.
 \item $\jac{m}$ has a conjugate pair of purely imaginary eigenvalues 
		$\lambda_1=\lambda_2^*=\ii w$. Solutions then 
		oscillate around \eqb{m} with period $2\pi/w$,
		and the equilibrium (called a \emph{center}) 
		is (neutrally) stable.
\end{itemize}

Taking into account equilibrium conditions $\dot{x}=\dotppar=0$, we find
from \refeq{eq:jacDef}
\[
 \jac{m} = 	\left(
		  \begin{array}{cc}
			0 		& 1/\gamma_m\\
			\jacElem{m}{2}{1} 	& 0
		  \end{array}
		\right),
\]
where
\[
  \jacElem{m}{2}{1}=\frac{1}{\gamma}
      \left[\frac{a^2_m\,(a'_m)^2}{\gamma^2_m}-(a'_m)^2+a^2_m\right]
				-\begin{cases}  n_0\,, &  x>0\,,\\
						  0\,, &  x<0\,.
                                 \end{cases}
\]
Here, we have defined $a_m=a(x_m)$, $a'_m=a'(x_m)$, 
$\gamma_m=\sqrt{1+a^2(x_m)}$,
and we have used \refeq{eq:m3}.

Eigenvalues of $\jac{m}$ are given by
\beq\label{eq:eilJac}
 \lambda_{1,2}(x_m)=\pm\sqrt{\jacElem{m}{2}{1}/\gamma_m}\,.
\eeq

In the vacuum, $x<0$, equilibria correspond to either $a(x_k^-)=0$ ($k$ even, nodes of 
the standing wave) or $a'(x_k^-)=0$ ($k$ odd, antinodes of the standing wave),
where the $x_k^-$ are given by \refeq{eq:eqbVac}. Then,
\refeq{eq:eilJac} yields
by using \refeqset{eq:eqbVac}{eq:a_csl},
\beq
 \lambda_{1,2}(x_k^-)=\pm\begin{cases}
		    \ii \sqrt{2}\,a_0\,, & k \text{ even,} \\
		    \frac{\sqrt{2}\,a_0}{\sqrt{1+2\,a_0^2}}\,, & k \text{ odd.} 
                  \end{cases}
\eeq
Thus, in the vacuum, equilibria alternate between being (neutrally) stable
($k$ even, nodes) and unstable ($k$ odd, antinodes). 

In the \csl, $x>0$, we have
\[
 \lambda_{1,2}(x_m) = \pm \frac{1}{\gamma_m^2}\sqrt{\gamma_m^2(a_m^2-(a'_m)^2)+a_m^2(a'_m)^2-\gamma_m^3 n_0}\,.
\]

For the equilibrium \eqb{b} at the plasma boundary, $x=x_b$, 
we get from \refeq{eq:a_csl}
\beq\label{eq:eilQb}
 \lambda_{1,2}(x_b) = \pm\sqrt{a_b^4 + 2 a_b^2 - 2 a_0^2-n_0(1+a_b^2)^{3/2}}/(1+a_b^2)\,.
\eeq
Linear (neutral) stability of \eqb{b} requires
\[
 a_b^4 + 2 a_b^2 - 2 a_0^2-n_0(1+a_b^2)^{3/2}<0\,,
\]
or, using 
\refeq{eq:ab} to eliminate $a_0$, 
\beq\label{eq:xbstab}
 2(a_b^2+n_0)-3n_0(1+a_b^2)^{1/2}<0\,.
\eeq
The same condition for linear stability of the equilibrium at $x_b$ was obtained
by Eremin\etal~\rf{eremin2010} by considering the infinitesimal variation in
electrostatic and ponderomotive force experienced by an electron whose position
has been perturbed infinitesimally to $x_b-|\delta x|$.
Condition \eqref{eq:xbstab} also coincides with condition \eqref{eq:stab_0} 
of existence of a stationary standing wave obtained by 
Cattani\etal\rf{cattani2000}. Therefore, as long as an equilibrium at $x_b$ exists,
it is neutrally stable. 

Assessing stability of the equilibria  with $0<x_1<x_b$ analytically
is somewhat more difficult [even when an explicit expression such as 
\refeq{eq:x1} is available]. 
We can, however, conclude that \eqb{1} is an unstable 
equilibrium on topological grounds. If we assume \eqb{1} to be stable, 
then motion in its neighborhood would be oscillatory. Therefore, 
a point $(x_s,0)$ in phase space with $x_1<x_s<x_b$ 
would be shared by oscillatory solutions encircling $\eqb{1}$ and $\eqb{b}$ (in phase space).
This would contradict uniqueness of solutions, unless 
the point $(x_s,0)$ were to be reached in infinite time, \ie\ unless it is an 
unstable equilibrium. 
However, by construction there is no equilibrium beween $Q_1$ and $Q_b$.
In fact, the 
degenerate oscillations introduced in this argument, which reach $Q_1$ in infinite time,
are the familiar separatrices of bounded and unbounded motion, 
which we will now study in detail.

\subsection{Separatrices\label{s:sep}}

In the vacuum ($x<0$), all unstable equilibria at $x_k^-$ (with $k$ odd) correspond 
to the same value of $H$,
\beq
  H(x_k^-,0) = \sqrt{1+2\,a_0^2}\,.
\eeq
Conservation of $H$, thus allows for a \emph{heteroclinic connection}, \ie, for an
orbit which starts infinitesimally close to \eqb{k} and ends infinitesimally close
to \eqb{k+2} or \eqb{k-2} (in infinite time). According to \refeq{eq:PSorb} these
orbits obey
\beq\label{eq:hetVac}
  \ppar(x)=\pm\sqrt{2a_0^2-a^2(x)}\,.
\eeq
Heteroclinic connections, \refeq{eq:hetVac}, act as separatrices of bounded
and unbounded motion, see \reffig{f:PS}.

Within the \csl\ ($0<x<x_b$), an unstable equilibrium, \eg\ \eqb{1} in \reffig{f:PS},
will in general have $H(x_1,0)\neq H(x_3,0)$ since $H$ now also includes an electrostatic
field contribution. Therefore, a heteroclinic connection from $\eqb{3}$ to \eqb{1} is
not possible, and the separatrix starting out at $\eqb{3}$ is a \emph{homoclinic}
connection, \ie\ an orbit that returns to \eqb{3} in infinite time. For the same
reason, the separatrix labeled $\mathrm{B}$ in \reffig{f:PS} starts in the neighborhood
of \eqb{1} and wanders off to $x=-\infty$, while the separatrix labeled $\mathrm{A}$
starts at $x=-\infty$ and ends at $\eqb{1}$.

Of greatest importance in the following discussion are the separatrices labeled $\Gamma$ and
$\Delta$, as they determine the region within which motion close to $\eqb{b}$
is oscillatory. The equations of the separatrices $\Gamma$ and
$\Delta$ are given by \refeq{eq:PSorb} with $(x_0,\pparo)=(x_1,0)$
[on separatrix $\Gamma$, motion is backwards in time and $(x_1,0)$ is a final, 
rather than initial, condition]. 
The point on separatrix $\Gamma$ at position $x_b$ (at the plasma boundary)
then defines a critical momentum $\pcr$,  
given by
\beq\label{eq:pcr}
    \pcr=-\left[\left[\sqrt{1+a^2(x_1)}+n_0(x_1^2-x_b^2)/2\right]^2-a_b^2-1\right]^{1/2}\,.
\eeq

If a single electron at the edge of the plasma $x_b$ is given an initial
momentum $-|\Delta p_x|$, with $|\Delta p_x|<|\pcr|$, it will move within the limits
set by separatrices $\Gamma$ and $\Delta$, returning back to the plasma. If, on
the other hand $|\Delta p_x|>|\pcr|$, the electron's motion will be unbounded
and it will escape to the vacuum. Alternatively, one can define a critical value of
the Hamiltonian
\beq\label{eq:Hcr}
  \Hcr\equiv H(x_1,0) = \sqrt{1+a^2(x_1)}+n_0\,x_1^2/2\,.
\eeq
Motion of electrons with $H(x_b,\ppar)>\Hcr$ and $\ppar<0$ will be unbounded.

\refEq{eq:pcr} shows that $|\pcr|$ is always non-zero as long as $x_1\neq x_b$;
for fixed $a_0$ it becomes smaller as $n_0$ decreases and $x_1$ approaches $x_b$, 
vanishing at the threshold $\nC$ given by \refeq{eq:stab_0}. This behavior 
is illustrated in 
\reffig{f:sep} for $a_0=7$. (See also \reffig{f:pxmin}.)

With the above results it becomes clear that finite perturbations 
of initial conditions of electrons at
the edge of the plasma, for example due to longitudinal electron heating, 
could lead to electrons escaping toward the vacuum even when $\eqb{b}$ is stable in
the linear approximation, provided that the perturbation (here negative momentum) 
is large enough. Our main conclusion is that
pulse propagation by expulsion of electrons toward the vacuum
could occur for densities higher than the threshold density 
$\nC$ predicted by the cold fluid approximation.
In \refsect{s:numerics} we show that electron heating at the edge of the plasma
indeed provides a mechanism by which electrons 
acquire sufficient momentum to escape toward the vacuum.

\section{PIC simulations\label{s:numerics}}

To investigate the transition from total reflection to \rsit, we perform
PIC simulations\rf{birdsall} using the one-dimensional in space, 
three-dimensional in velocity (1D3V) code \squash\rf{grech2020}. The code uses the 
finite-difference, time-domain approach for solving Maxwell's equations\rf{taflove},
and the standard (Boris) leap-frog scheme for solving the macro-particle equations of
motion\rf{boris1970}. Charge conservation is ensured by using the method proposed by
Esirkepov when projecting the currents\rf{esirkepov_CPC_2001}.

In all simulations presented here, ions are immobile and only electron motion is
considered. We use the spatial resolution $dx=\lambda_L/500$ and time step 
$dt=\tau_L/1000$, where $\lambda_L$ and $\tau_L$ are the laser wavelength and
duration of one optical cycle, respectively. 
Up to 1000 macro-particles per cell have been used.

The plasma extends from $x=0$ to $x=L_p$, with a
constant initial density $n_0$ and 
electron temperature $T_0\simeq5\cdot10^{-4}$ (in units
of $m_e c^2$). The plasma 
size $L_p$ is chosen so that $L_p>c\,\tint$, 
where \tint\ is the laser-plasma interaction time. 
Hence, the plasma is long enough to be considered semi-infinite. 
The CP laser pulse [as described by 
\refeq{eq:pulse}] is incident from $x<0$ onto the plasma. In this work
we consider laser field amplitudes in the range $a_0=1-30$. The laser pulse
profile is trapezoidal, \ie\ the intensity increases linearly within a
rise time $\tr$, up to a maximum value $a_0^2/2$, and 
we consider the exemplary cases $\tr=0.25\,\tau_L$ and $\tr=4\,\tau_L$. 

\begin{figure}[htpb]
  \begin{center}
    \includegraphics[width=\cwidth, clip=true]{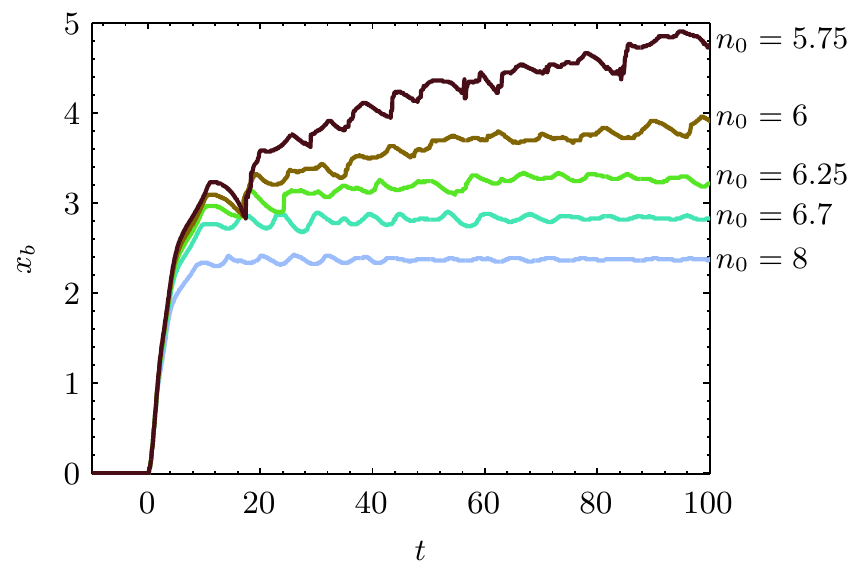}
  \end{center}
  \caption{\cweb\ Position of the maximum value of the electrostatic field $\xEmax$ as
  a function of time from PIC simulations with different densities 
  and $a_0=15$, $\tr=0.25\,\tl$. 
  }
  \label{f:xb_osc}
\end{figure}

\refFig{f:threshold_geom}(a) summarizes our findings on \rsit, comparing the threshold 
density $\nC(a_0)$ predicted by Cattani\etal\rf{cattani2000} with our 1D3V PIC
simulation results. In order to determine whether \rsit\ occurs or not 
in a simulation, the position $x_b$ of the maximum electrostatic field is  
plotted as a function of time (see \reffig{f:xb_osc}).
The regime of total reflection is characterized by the formation of a \csl\
with (approximately) constant thickness $x_b$ (\reffig{f:xb_osc}, $n_0=6.7-8$).
On the other hand, \rsit\ is  associated with front propagation at 
an approximately constant 
velocity $\vf$, so that the position of the maximum electrostatic 
field increases linearly with time (\reffig{f:xb_osc}, $n_0=5.75-6$).
This allows us to place lower and upper bounds on \rsit\ threshold density, 
for a certain $a_0$, indicated by error bars in \reffig{f:threshold_geom}. 
For densities within these limits, it is hard to decide whether \rsit\ occurs or not
(\reffig{f:xb_osc}, $n_0=6.25$).
%
%

In the next subsections we examine in detail typical cases of total reflection and 
front penetration.

\subsection{Total reflection}
 
\begin{figure}[htpb]
  \begin{center}
    \includegraphics[width=\cwidth, clip=true]{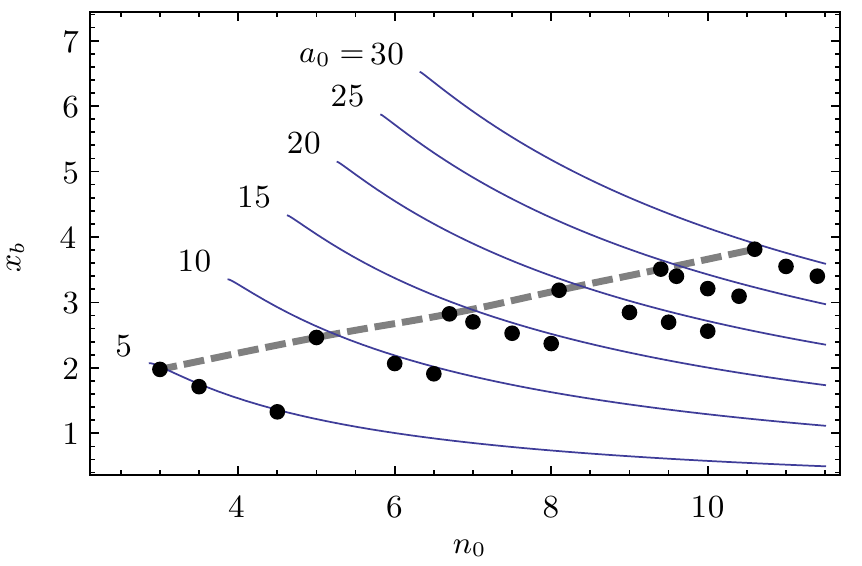}
  \end{center}
  \caption{\cweb\ Comparison of cold-fluid model prediction for $x_b$ 
  (blue, solid lines) with
  the (time-averaged) position of the maximum electrostatic field in our PIC simulations (dots), 
  with $\tr=0.25\,\tl$. 
  For values of $n_0$
  to the left of the thick, gray, dashed line RSIT occurs and 
  $x_b$ does not reach a constant average value in our PIC simulations.
  }
  \label{f:xb}
\end{figure}

\begin{figure}[htpb]
  \begin{center}
    \includegraphics[width=\cwidth, clip=true]{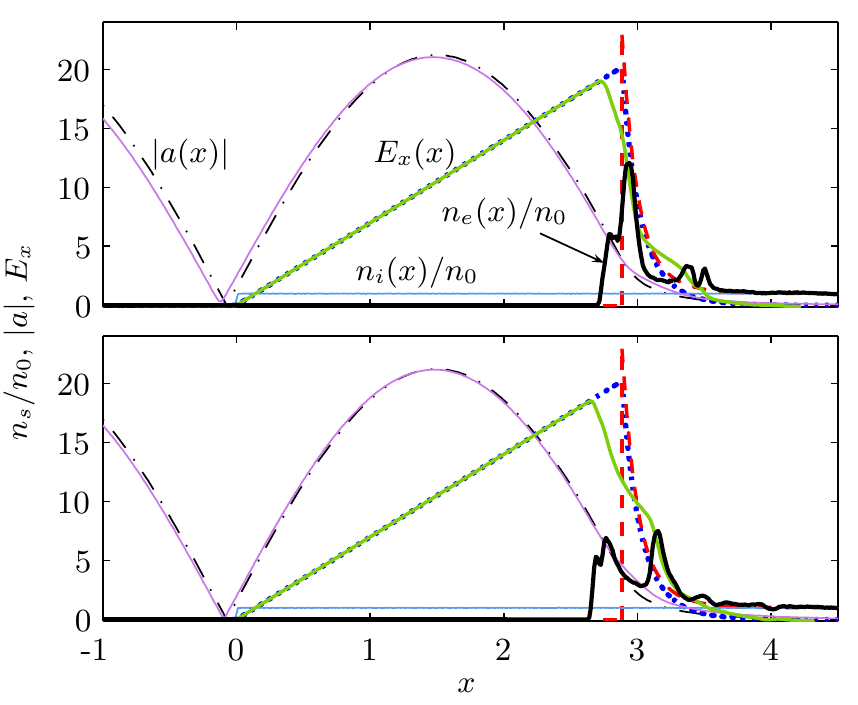}
  \end{center}
  \caption{\cweb\ Electron density and field profiles
    from PIC simulations for $a_0=15$, $n_0=7$
    at $t=2.55\,\tl$ (top panel) and $t=2.95\,\tl$ (bottom panel). 
    The stationary cold-fluid model solution for the electron density 
    (red, dashed line), electrostatic field (blue, dotted line) and
	vector potential envelope (black, dash-dotted line)
    are also shown. Note that densities have been rescaled to the unperturbed
	density $n_0$ for better readability.
  }
  \label{f:profComp}
\end{figure}

\begin{figure*}
  \begin{center}
    \includegraphics[width=\textwidth, clip=true]{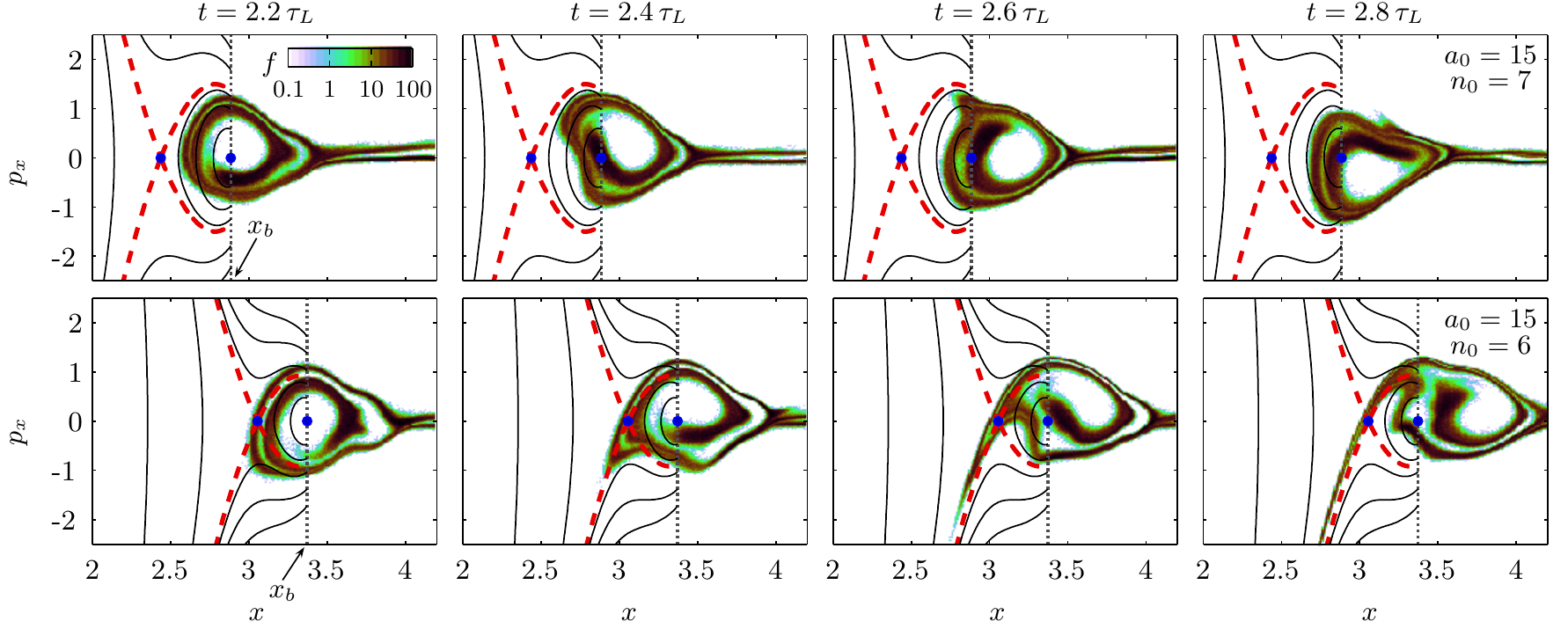}
  \end{center}
  \caption{\cweb\ Comparison between phase space separatrices as predicted by the stationary,
    cold-fluid model, and  single particle distribution function $f(x,\ppar)$ from PIC simulation results for $a_0=15$ and $n_0=7$ (top
    row), $n_0=6$ (bottom row) and rise time $\tr=0.25\,\tl$. 
	Snapshots are shown $0.2\,\tl$ apart.
    The plasma boundary ($x=x_b$), as predicted by the cold-fluid model, is 	
    indicated by a black, dotted, vertical line. The color coding of trajectories
    follows \reffig{f:PS}. Note the logarithmic scale in the color coding of $f(x,p_x)$.
  }
  \label{f:PSpic}
\end{figure*}

Whenever total reflection occurs, the system 
eventually settles to a quasi-stationary state. 
The size of the charge separation layer $x_b$ 
remains constant or slightly oscillatory around
a value that is found to be in good agreement with
the theoretical prediction of the cold-fluid model [\refeq{eq:xb}], see \reffig{f:xb}. 
The same is
true for the field and density profiles;
a worst case agreement is shown in \reffig{f:profComp},
where the quasistationary state 
reached for $a_0=15$, $n_0=7$ and $\tr=0.25\tau_L$ is 
close to the numerical \rsit\ threshold (the agreement becomes better
for higher $n_0$ or larger $\tr$). 
Although the density profile presents oscillations, 
the fields in the \csl\ and vacuum
 agree very
well with the predictions of cold-fluid theory. 
This justifies \emph{a posteriory} our use 
of stationary cold fluid theory predictions
for the fields in the vacuum to analyze single electron
phase space in \refsect{s:PS}. 
The phase portrait for $a_0=15$, $n_0=7$ and $\tr=0.25\tau_L$
is shown in the top row of \reffig{f:PSpic}. 
It is clearly seen that electrons in the \cel\ do not
have zero longitudinal momentum $p_x$ as the stationary cold-fluid 
model suggests, but rather oscillate
around $x_b$ [the latter being in good agreement with \refeq{eq:xb}]. 
As the minimum momentum attained by electrons, which we will call $\pth$, 
is smaller in absolute value than the critical momentum required to move 
beyond the limits 
set by the separatrices of bounded and unbounded motion, $|\pth|<|\pcr|$, 
electrons which cross the plasma boundary $x_b$ do not escape into the vacuum but
rather re-enter the \cel.
%
%

%
%

%

\begin{figure}
  \begin{center}
    \includegraphics[width=\cwidth, clip=true]{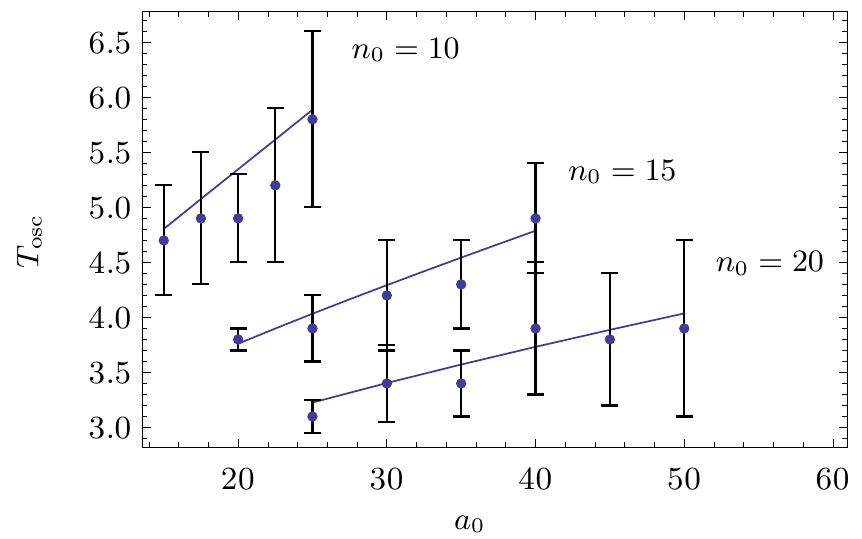}
  \end{center}
  \caption{\cweb\ Period of oscillations of $x_b$ as a function of $a_0$ for three different
	$n_0$. Points with error bars correspond to the periods as deduced from our PIC simulations,
	while the solid lines correspond to $T_{\mathrm{osc}}=1.75\, T_{Q_b}=1.75\times 2\pi/\Im\lambda_1$,
	where $\lambda_1$ is given by \refeq{eq:eilQb}.
  }
  \label{f:xb_osc_T}
\end{figure}

As can be seen in \reffig{f:xb_osc}, for $n_0=6.7-8$, the position of 
the plasma boundary $x_b$ oscillates in time, leading to oscillations of 
the maximum electrostatic field. These oscillations 
can be related to the excursion of electrons in the region $x<x_b$,
\cf\ the top panel of \refFig{f:PSpic}.
To verify this, we plot in \reffig{f:xb_osc_T} the period $T_{\mathrm{osc}}$ 
of these oscillations for different $a_0$ and $n_0$ well in 
the regime of total reflection. 
The frequency of these oscillations is not linked to the plasma frequency
(observe the dependence on $a_0$ in \reffig{f:xb_osc_T}) but rather on the 
frequency of oscillations of electrons around the equilibrium $Q_b$. 
If we ignore the role of the self-consistent fields within the plasma,
the characteristic period of oscillation in the linear neighborhood of $Q_b$ 
reads $T_{Q_b}=2\pi/\Im\lambda_1$,
where $\lambda_1$ is the eigenvalue given by \refeq{eq:eilQb}. 
As shown in \reffig{f:xb_osc_T},
we find $T_{\mathrm{osc}}\propto T_{Q_b}$.
We also note the similarity
of these oscillations with the so-called 
piston oscillations in laser hole-boring\rf{Schlegel_POP_2009}, although
in the present case the oscillations only involve electrons.

%
%

\subsection{\rsit}

\begin{figure*}
  \begin{center}
    \includegraphics[width=\textwidth, clip=true]{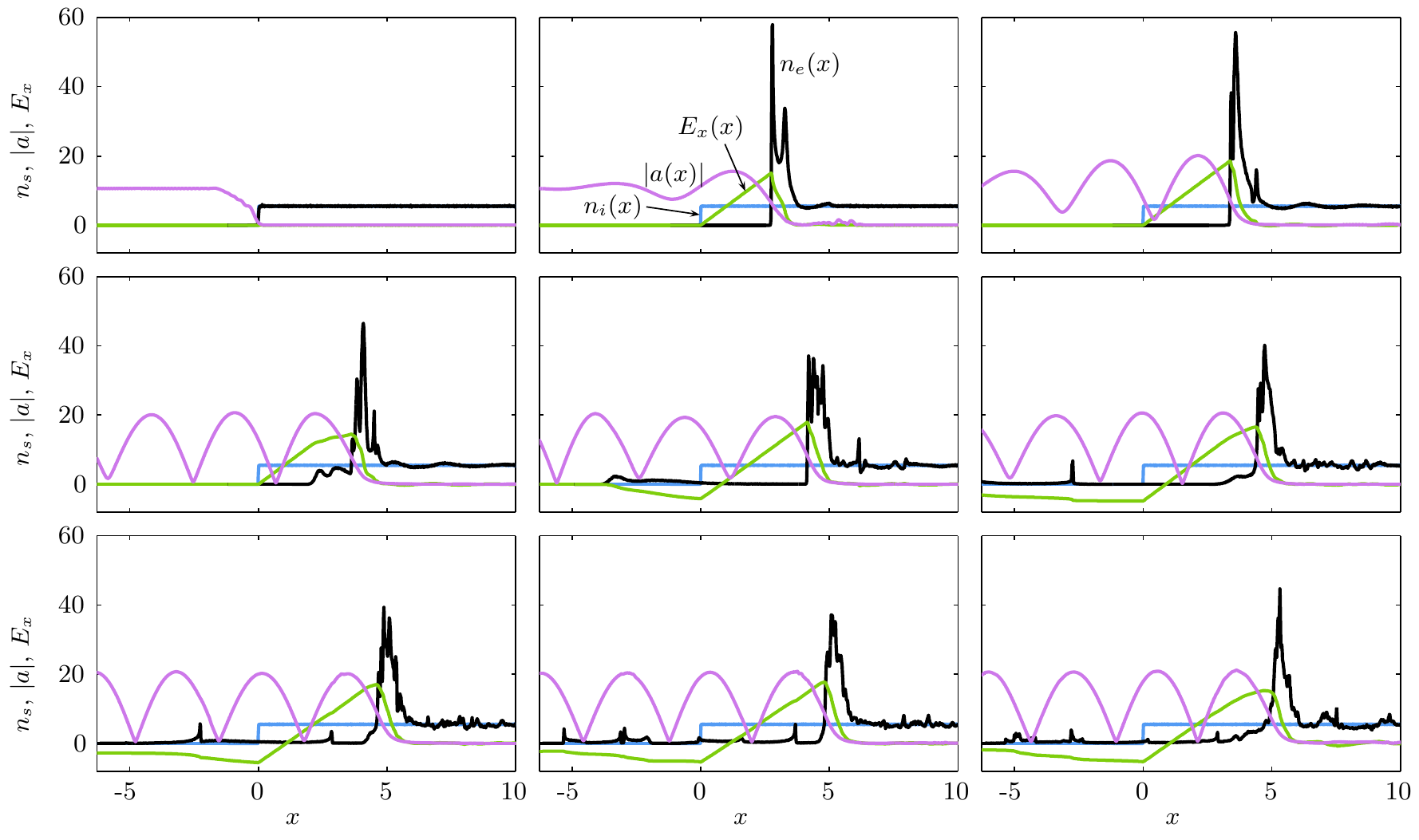}
  \end{center}
  \caption{\cweb\ Field and density evolution for \rsit\ above the 
    cold-fluid theory threshold, $a_0=15$, $n_0=5.5$. Snapshots are taken
	one laser period apart, starting at $t=0$.}
	\label{f:PICprop}
\end{figure*}

The cold fluid model presented in \refsect{s:review} predicts a sharp threshold, 
either for density $n_0$ or laser amplitude $a_0$, for \rsit. 
However, as already mentioned above, one of the main results of this paper 
is that our PIC simulations clearly show \rsit\ in a parameter region 
where the cold fluid model predicts total reflection 
[area (B) in \reffig{f:threshold_geom}(a)]. A typical case of \rsit\ in 
this regime is presented in \reffig{f:PICprop}, 
where $a_0=15$, $n_0=5.5$  and $\tr=0.25\tau_L$. 
Charge separation and compressed electron layers are formed in the 
early stages of interaction, with profiles that 
agree well with the predictions of cold-fluid theory.
However, electrons escape the \cel, and the pulse can propagate (see middle row
of \reffig{f:PICprop}). 
The mechanism of propagation is rather complex,
but its initial phase can be
intuitively understood as follows. When a sufficiently high number of electrons 
escapes from the \cel\ to the vacuum, the electrostatic field 
within the \csl\ decreases,
the ponderomotive force is no longer balanced and the laser pulse
can push the \cel\ deeper into the plasma. The increase of the \csl\ size tends
to compensate the force imbalance, but as more and more electrons escape, 
the pulse continues
to propagate deeper into the plasma. We note that once electrons escape
and propagation commences the stationary model is no longer valid and electron
dynamics becomes complex, with electron bunches leaving and re-entering 
the plasma (see \reffig{f:PICprop} and \refref{eremin2010}).

To understand how the shrinking of the width of separatrices in phase space with
decreasing density (and constant $a_0$) leads to propagation, we examine 
the phase space portrait for $a_0=15$, $n_0=6.0$ and $\tr=0.25\,\tl$, 
which corresponds to a case just below the numerical density threshold for \rsit,
see the bottom row of \reffig{f:PSpic}. 
In this case, the minimum momentum acquired by electrons in the 
\cel\ satisfies $|\pth|>|\pcr|$ and electrons move outside 
the separatrix of bounded and unbounded motion, eventually reaching the vacuum,
while the \cel\ moves deeper into the plasma. 
\begin{figure}[htpb]
  \begin{center}
    \includegraphics[width=\cwidth, clip=true]{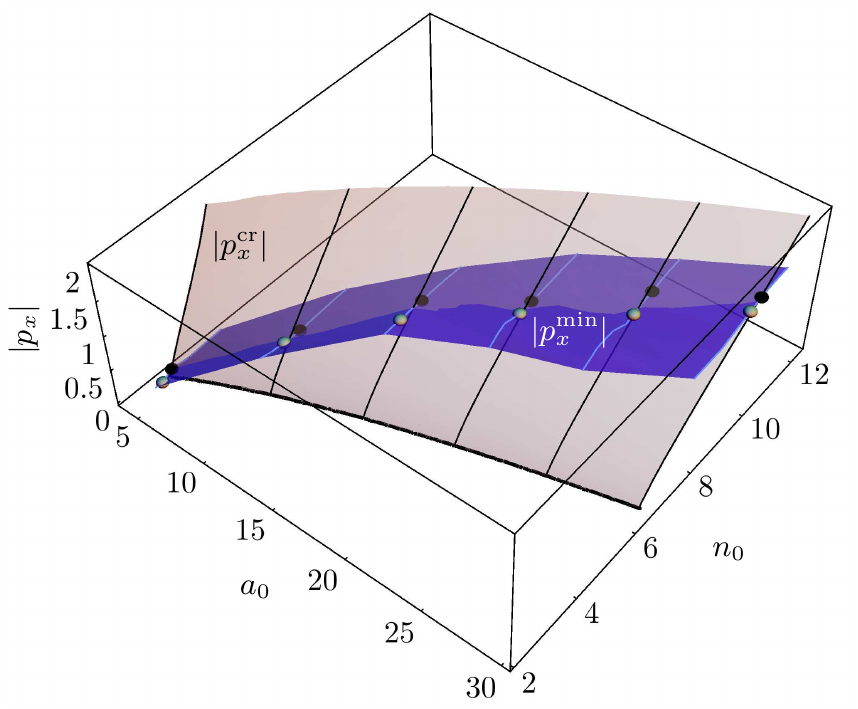}
  \end{center}
  \caption{\cweb\ The absolute value of momentum 
	  $|\pcr|$ corresponding to the separatrix
	  of bounded and unbounded motion for electrons
	  at $x_b$ for different $a_0$ and $n_0$, according to \refeq{eq:pcr} is shown as a light-gray surface.
	  An estimate of the absolute value of the minimum 
	  momentum $|\pth|$ attained by electrons in the \cel, as determined
	  by our PIC simulations with $\tr=0.25\tl$, is shown as a dark-blue surface. The 
	  light- and dark-colored points represent 
	  PIC simulation results corresponding to onset of \rsit\ 
	  and total reflection, respectively, for $a_0=5,\,10,\ldots,30$.
	  On both surfaces, lines of constant $a_0$ are drawn to guide the eye.
	  The contour $|\pcr|=0$ (black, thick, solid line) 
	  corresponds to the threshold for \rsit\ predicted by cold-fluid
	  theory.
	  }
	\label{f:pxmin}
\end{figure}

\refFig{f:pxmin} provides a further verification of the role the
longitudinal electron heating plays in enabling electrons
to escape from the \cel\ into the vacuum. 
We use \refeq{eq:pcr} to plot $|\pcr|$ as a function of $a_0$ and $n_0$ 
(light-gray surface). 
For a given rise time, here $\tr=0.25\,\tl$, we also plot, as a 
function of $a_0$ and $n_0$, the absolute value of the 
minimum momentum $|\pth|$ acquired 
by electrons in the \cel\ as inferred from our PIC simulations (dark-blue surface).
To reduce noise we average $|\pth|$ 
over one laser period (starting at $t\simeq2\tl$),  
or at most until electrons escape. Thus, our $|\pth|$ is generally
slightly underestimated, however the intersection of the two surfaces $|\pcr|$
and $|\pth|$ lies within the limits set by the error bars in \reffig{f:threshold_geom}(a).
Note that $|\pth|$ is getting smaller with decreasing $a_0$, and one 
recovers the threshold predicted by cold-fluid theory for $a_0\lesssim5$, 
where the longitudinal electron momenta become negligible [compare with
\reffig{f:threshold_geom}(a)].

\subsection{Dependence on rise time}

\begin{figure}[htpb]
 \begin{center}
  \includegraphics[width=\cwidth,clip=true]{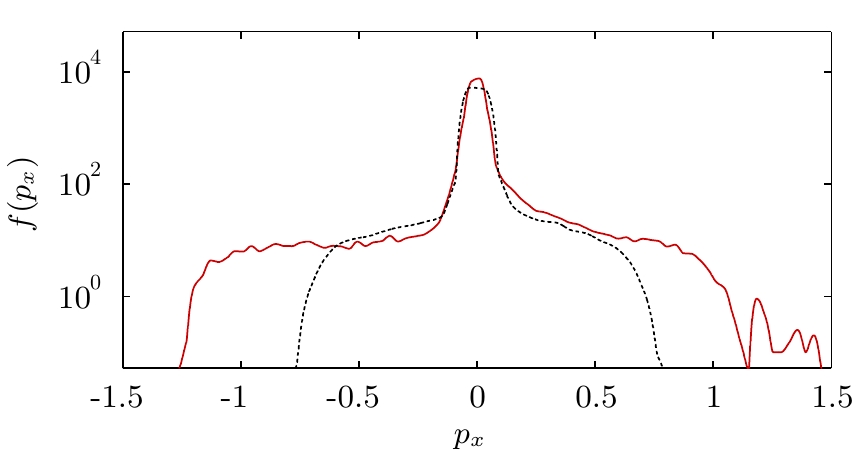}
  \caption{
  \cweb\ Space-integrated (over all $x$) longitudinal momentum 
  distribution $f(\ppar)$ at $t=15\,\tl$ for
  $a_0=15$, $n_0=7$ and $\tr=0.25\,\tl$ (red, solid curve), $\tr=4\,\tl$
  (black, dotted curve). Both cases correspond to total reflection. The central
  peak corresponds to the plasma bulk.
  }
  \label{f:fpx_comp}
 \end{center}
\end{figure}

As we have seen, the threshold for transition between total reflection and
\rsit\ clearly depends on the longitudinal momenta of the electrons 
in the \cel. As these momenta come from collisionless heating of the electrons, 
we may expect  that the \rsit\ threshold also depends on the laser pulse profile.
As can be seen in \reffig{f:threshold_geom}(a), the deviation of the
numerically obtained \rsit\ threshold
from the predictions of cold fluid theory is smaller for a pulse with larger rise time, 
suggesting a less significant electron heating in the CEL at given $a_0$ and $n_0$.
The effect of pulse rise time on the width of the
longitudinal electron momentum distribution function
is shown in \reffig{f:fpx_comp}, where the space-integrated distribution for $a_0=15$, $n_0=7$ is compared 
for the cases $\tr=0.25\tl$ and $\tr=4\tl$. 
The stiffer pulse clearly corresponds to a larger $|\pth|$.
%

\begin{figure}[htpb]
  \begin{center}
    \includegraphics[width=\cwidth, clip=true]{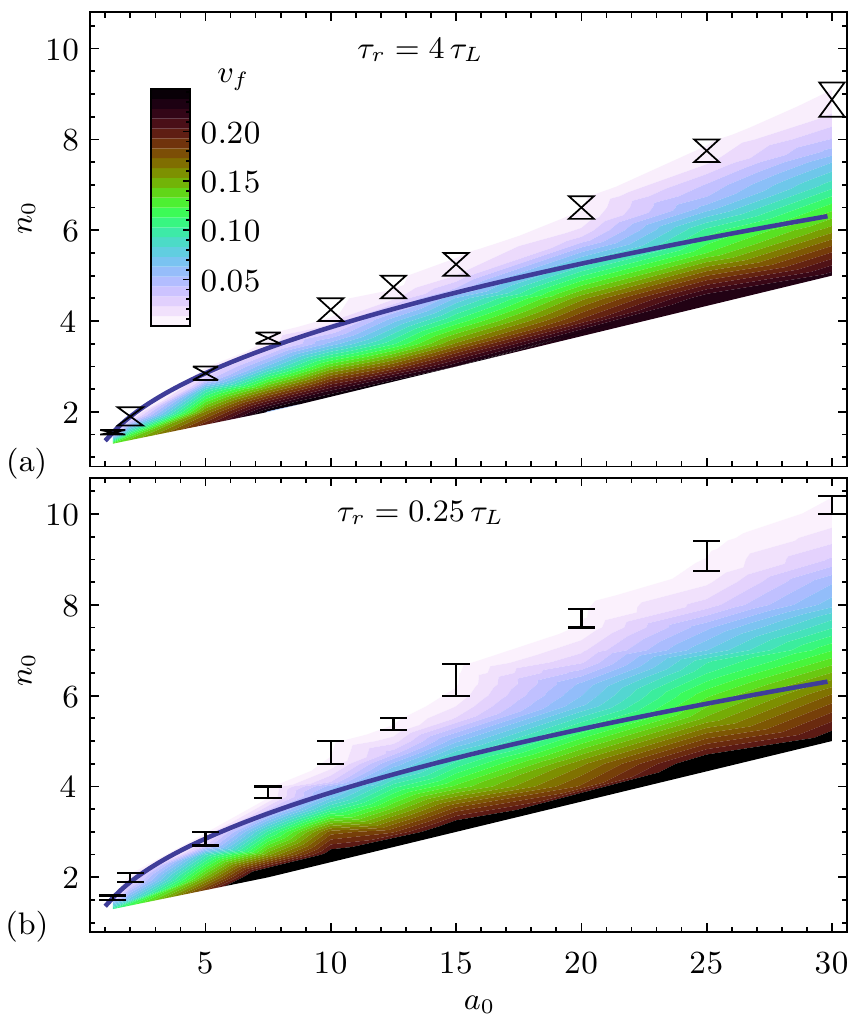}
  \end{center}
  \caption{\cweb\ Front velocity $v_f$ as measured from PIC simulations with two different
      pulse rise times, (a) $\tr=4\,\tl$, (b) $\tr=0.25\,\tl$. 
       The blue, solid line and error bars are the same as described in the caption of \reffig{f:threshold_geom}(a).
      The lower density
      range of simulations performed for a given $a_0$ has been set to improve
      readability.
    }
  \label{f:vf}
\end{figure}

In \reffig{f:vf}, we moreover compare the 
front propagation
velocity, $v_f$, for two sets of simulations with rise 
times $\tr=0.25\tl$ and $\tr=4\tl$.
The front propagation speed $v_f$ is determined by the slope of the curves 
$\xEmax(t)$, see \reffig{f:xb_osc}. We have studied cases of
propagation for different $a_0$ and $n_0$ close to the threshold predicted 
by cold-fluid theory, for which $v_f$ ranges from $10^{-3}\,c$ up to $0.25\,c$,
see \reffig{f:vf}. 
Within the error bars for the transparency threshold,
$v_f$ takes values too small to reliably indicate propagation 
(\ie\ beyond the accuracy permitted by our spatial and temporal resolution).
As \reffig{f:vf} shows, the propagation velocity $v_f$ for the 
same $a_0$ and $n_0$ is generally lower for the pulse with the 
larger rise time, $\tr=4\,\tl$. Nevertheless, 
for higher $a_0$, $v_f$ is far from negligible 
for densities lying well above the cold-fluid threshold,
even for the case with larger rise time.
%
%

\section{Discussion and conclusions\label{s:disc}}

The relativistic, cold-fluid, 
stationary solutions of \refrefs{marburger1975,cattani2000,goloviznin2000}
provide a convenient starting point to investigate the threshold of \rsit,
even in the presence of longitudinal electron heating. 
While the fields inside the plasma clearly differ from the predictions of cold
fluid theory, the fields in the \csl\ and vacuum are
rather insensitive to density fluctuations within the plasma. 
Therefore, the dynamics of a test electron in the \csl\ or the vacuum
can be accurately described using the fields of the stationary problem. 
This finding allows us to specify separatrices of bounded and unbounded 
motion for single electron dynamics, encapsulating the competition of ponderomotive
and electrostatic forces at the edge of the plasma. 

We have shown that one can define a critical momentum $|\pcr|$, \refeq{eq:pcr},
or value of the Hamiltonian \Hcr, \refeq{eq:Hcr}, corresponding to
the separatrix which delimits oscillatory motion around the equilibrium position
at the edge of the \cel. 
When a sufficiently high number of electrons at the edge of the \cel\ have $|\ppar|>|\pcr|$
and escape to the vacuum, \rsit\ occurs.
In this work, we did not focus on the mechanism that provides
momentum to electrons, \ie\  
we did not attempt to provide a model for the collisionless 
heating mechanism.
We did however show, through our numerical study  
of  the impact of the pulse rise time, that the pulse shape crucially affects
longitudinal heating and that stronger heating results in a higher 
threshold density for \rsit.  A detailed model for electron heating, 
which would allow us to predict $|\pth|$ 
rather than infer it from PIC simulations, as done in \reffig{f:pxmin}, 
will be pursued elsewhere. We stress that, although in more realistic scenarios
of laser-plasma interaction the actual heating at
the plasma boundary would depend on several factors (see \refref{sanz2012} for a recent study), 
the basic mechanism of electron escape into the vacuum at high enough momentum is expected to be
the same. 

In summary, we have used a dynamical systems approach to bridge the cold-fluid and kinetic
levels of \rsit\ description. Deviations of PIC simulations from
cold-fluid theory predictions are explained as a longitudinal heating effect
induced by the incident laser pulse. 
The pulse temporal profile clearly affects electron heating and through it the 
threshold of \rsit. While there are several experimental works addressing \rsit\ 
in the case of linearly polarized laser pulses~\rf{giulietti1997, fuchs1998,willingale2009,palaniyappan2012}, 
to the best of our knowledge the verification of \rsit\ for CP light remains elusive. 
We hope that our results trigger further 
investigations in this domain, as the reported dependency of the \rsit\ threshold 
on the pulse profile could provide a versatile tool for high-contrast 
CP laser pulse characterization.

\section{Acknowledgments}

We would like to thank A. Debayle, L. Gremillet, A. Macchi and A. Pukhov 
for helpful comments.


\pagebreak

%



\begin{thebibliography}{37}%
\makeatletter
\providecommand \@ifxundefined [1]{%
 \@ifx{#1\undefined}
}%
\providecommand \@ifnum [1]{%
 \ifnum #1\expandafter \@firstoftwo
 \else \expandafter \@secondoftwo
 \fi
}%
\providecommand \@ifx [1]{%
 \ifx #1\expandafter \@firstoftwo
 \else \expandafter \@secondoftwo
 \fi
}%
\providecommand \natexlab [1]{#1}%
\providecommand \enquote  [1]{``#1''}%
\providecommand \bibnamefont  [1]{#1}%
\providecommand \bibfnamefont [1]{#1}%
\providecommand \citenamefont [1]{#1}%
\providecommand \href@noop [0]{\@secondoftwo}%
\providecommand \href [0]{\begingroup \@sanitize@url \@href}%
\providecommand \@href[1]{\@@startlink{#1}\@@href}%
\providecommand \@@href[1]{\endgroup#1\@@endlink}%
\providecommand \@sanitize@url [0]{\catcode `\\12\catcode `\$12\catcode
  `\&12\catcode `\#12\catcode `\^12\catcode `\_12\catcode `\%12\relax}%
\providecommand \@@startlink[1]{}%
\providecommand \@@endlink[0]{}%
\providecommand \url  [0]{\begingroup\@sanitize@url \@url }%
\providecommand \@url [1]{\endgroup\@href {#1}{\urlprefix }}%
\providecommand \urlprefix  [0]{URL }%
\providecommand \Eprint [0]{\href }%
\providecommand \doibase [0]{http://dx.doi.org/}%
\providecommand \selectlanguage [0]{\@gobble}%
\providecommand \bibinfo  [0]{\@secondoftwo}%
\providecommand \bibfield  [0]{\@secondoftwo}%
\providecommand \translation [1]{[#1]}%
\providecommand \BibitemOpen [0]{}%
\providecommand \bibitemStop [0]{}%
\providecommand \bibitemNoStop [0]{.\EOS\space}%
\providecommand \EOS [0]{\spacefactor3000\relax}%
\providecommand \BibitemShut  [1]{\csname bibitem#1\endcsname}%
\let\auto@bib@innerbib\@empty
\bibitem [{Note1()}]{Note1}%
  \BibitemOpen
  \bibinfo {note} {The definition of the vector potential is given in Eq.~(\ref
  {eq:pulse})}\BibitemShut {NoStop}%
\bibitem [{\citenamefont {Akhiezer}\ and\ \citenamefont
  {Polovin}(1956)}]{akhiezer1956}%
  \BibitemOpen
  \bibfield  {author} {\bibinfo {author} {\bibfnamefont {A.~I.}\ \bibnamefont
  {Akhiezer}}\ and\ \bibinfo {author} {\bibfnamefont {R.~V.}\ \bibnamefont
  {Polovin}},\ }\href@noop {} {\bibfield  {journal} {\bibinfo  {journal} {Sov.
  Phys. JETP}\ }\textbf {\bibinfo {volume} {3}},\ \bibinfo {pages} {915}
  (\bibinfo {year} {1956})}\BibitemShut {NoStop}%
\bibitem [{\citenamefont {Kaw}\ and\ \citenamefont {Dawson}(1970)}]{kaw1970}%
  \BibitemOpen
  \bibfield  {author} {\bibinfo {author} {\bibfnamefont {P.}~\bibnamefont
  {Kaw}}\ and\ \bibinfo {author} {\bibfnamefont {J.}~\bibnamefont {Dawson}},\
  }\href {\doibase doi:10.1063/1.1692942} {\bibfield  {journal} {\bibinfo
  {journal} {Phys. Fluids}\ }\textbf {\bibinfo {volume} {13}},\ \bibinfo
  {pages} {472} (\bibinfo {year} {1970})}\BibitemShut {NoStop}%
\bibitem [{\citenamefont {Klimo}\ \emph {et~al.}(2008)\citenamefont {Klimo},
  \citenamefont {Psikal}, \citenamefont {Limpouch},\ and\ \citenamefont
  {Tikhonchuk}}]{klimo_PRSTAB_2008}%
  \BibitemOpen
  \bibfield  {author} {\bibinfo {author} {\bibfnamefont {O.}~\bibnamefont
  {Klimo}}, \bibinfo {author} {\bibfnamefont {J.}~\bibnamefont {Psikal}},
  \bibinfo {author} {\bibfnamefont {J.}~\bibnamefont {Limpouch}}, \ and\
  \bibinfo {author} {\bibfnamefont {V.~T.}\ \bibnamefont {Tikhonchuk}},\ }\href
  {\doibase 10.1103/PhysRevSTAB.11.031301} {\bibfield  {journal} {\bibinfo
  {journal} {Phys. Rev. ST Accel. Beams}\ }\textbf {\bibinfo {volume} {11}},\
  \bibinfo {pages} {031301} (\bibinfo {year} {2008})}\BibitemShut {NoStop}%
\bibitem [{\citenamefont {Robinson}\ \emph {et~al.}(2008)\citenamefont
  {Robinson}, \citenamefont {Zepf}, \citenamefont {Kar}, \citenamefont
  {Evans},\ and\ \citenamefont {Bellei}}]{Robinson_NJP_2008}%
  \BibitemOpen
  \bibfield  {author} {\bibinfo {author} {\bibfnamefont {A.~P.~L.}\
  \bibnamefont {Robinson}}, \bibinfo {author} {\bibfnamefont {M.}~\bibnamefont
  {Zepf}}, \bibinfo {author} {\bibfnamefont {S.}~\bibnamefont {Kar}}, \bibinfo
  {author} {\bibfnamefont {R.~G.}\ \bibnamefont {Evans}}, \ and\ \bibinfo
  {author} {\bibfnamefont {C.}~\bibnamefont {Bellei}},\ }\href
  {http://stacks.iop.org/1367-2630/10/i=1/a=013021} {\bibfield  {journal}
  {\bibinfo  {journal} {New J. Phys.}\ }\textbf {\bibinfo {volume} {10}},\
  \bibinfo {pages} {013021} (\bibinfo {year} {2008})}\BibitemShut {NoStop}%
\bibitem [{\citenamefont {Yan}\ \emph {et~al.}(2008)\citenamefont {Yan},
  \citenamefont {Lin}, \citenamefont {Sheng}, \citenamefont {Guo},
  \citenamefont {Liu}, \citenamefont {Lu}, \citenamefont {Fang},\ and\
  \citenamefont {Chen}}]{Yan_PRL_2008}%
  \BibitemOpen
  \bibfield  {author} {\bibinfo {author} {\bibfnamefont {X.~Q.}\ \bibnamefont
  {Yan}}, \bibinfo {author} {\bibfnamefont {C.}~\bibnamefont {Lin}}, \bibinfo
  {author} {\bibfnamefont {Z.~M.}\ \bibnamefont {Sheng}}, \bibinfo {author}
  {\bibfnamefont {Z.~Y.}\ \bibnamefont {Guo}}, \bibinfo {author} {\bibfnamefont
  {B.~C.}\ \bibnamefont {Liu}}, \bibinfo {author} {\bibfnamefont {Y.~R.}\
  \bibnamefont {Lu}}, \bibinfo {author} {\bibfnamefont {J.~X.}\ \bibnamefont
  {Fang}}, \ and\ \bibinfo {author} {\bibfnamefont {J.~E.}\ \bibnamefont
  {Chen}},\ }\href {\doibase 10.1103/PhysRevLett.100.135003} {\bibfield
  {journal} {\bibinfo  {journal} {Phys. Rev. Lett.}\ }\textbf {\bibinfo
  {volume} {100}},\ \bibinfo {pages} {135003} (\bibinfo {year}
  {2008})}\BibitemShut {NoStop}%
\bibitem [{\citenamefont {Macchi}\ \emph {et~al.}(2009)\citenamefont {Macchi},
  \citenamefont {Veghini},\ and\ \citenamefont {Pegoraro}}]{Macchi_PRL_2009}%
  \BibitemOpen
  \bibfield  {author} {\bibinfo {author} {\bibfnamefont {A.}~\bibnamefont
  {Macchi}}, \bibinfo {author} {\bibfnamefont {S.}~\bibnamefont {Veghini}}, \
  and\ \bibinfo {author} {\bibfnamefont {F.}~\bibnamefont {Pegoraro}},\ }\href
  {\doibase 10.1103/PhysRevLett.103.085003} {\bibfield  {journal} {\bibinfo
  {journal} {Phys. Rev. Lett.}\ }\textbf {\bibinfo {volume} {103}},\ \bibinfo
  {pages} {085003} (\bibinfo {year} {2009})}\BibitemShut {NoStop}%
\bibitem [{\citenamefont {Grech}\ \emph {et~al.}(2011)\citenamefont {Grech},
  \citenamefont {Skupin}, \citenamefont {Diaw}, \citenamefont {Schlegel},\ and\
  \citenamefont {Tikhonchuk}}]{Grech_NJP_2011}%
  \BibitemOpen
  \bibfield  {author} {\bibinfo {author} {\bibfnamefont {M.}~\bibnamefont
  {Grech}}, \bibinfo {author} {\bibfnamefont {S.}~\bibnamefont {Skupin}},
  \bibinfo {author} {\bibfnamefont {A.}~\bibnamefont {Diaw}}, \bibinfo {author}
  {\bibfnamefont {T.}~\bibnamefont {Schlegel}}, \ and\ \bibinfo {author}
  {\bibfnamefont {V.~T.}\ \bibnamefont {Tikhonchuk}},\ }\href
  {http://stacks.iop.org/1367-2630/13/i=12/a=123003} {\bibfield  {journal}
  {\bibinfo  {journal} {New J. Phys.}\ }\textbf {\bibinfo {volume} {13}},\
  \bibinfo {pages} {123003} (\bibinfo {year} {2011})}\BibitemShut {NoStop}%
\bibitem [{\citenamefont {Naumova}\ \emph {et~al.}(2009)\citenamefont
  {Naumova}, \citenamefont {Schlegel}, \citenamefont {Tikhonchuk},
  \citenamefont {Labaune}, \citenamefont {Sokolov},\ and\ \citenamefont
  {Mourou}}]{Naumova_PRL_2009}%
  \BibitemOpen
  \bibfield  {author} {\bibinfo {author} {\bibfnamefont {N.}~\bibnamefont
  {Naumova}}, \bibinfo {author} {\bibfnamefont {T.}~\bibnamefont {Schlegel}},
  \bibinfo {author} {\bibfnamefont {V.~T.}\ \bibnamefont {Tikhonchuk}},
  \bibinfo {author} {\bibfnamefont {C.}~\bibnamefont {Labaune}}, \bibinfo
  {author} {\bibfnamefont {I.~V.}\ \bibnamefont {Sokolov}}, \ and\ \bibinfo
  {author} {\bibfnamefont {G.}~\bibnamefont {Mourou}},\ }\href {\doibase
  10.1103/PhysRevLett.102.025002} {\bibfield  {journal} {\bibinfo  {journal}
  {Phys. Rev. Lett.}\ }\textbf {\bibinfo {volume} {102}},\ \bibinfo {pages}
  {025002} (\bibinfo {year} {2009})}\BibitemShut {NoStop}%
\bibitem [{\citenamefont {Schlegel}\ \emph {et~al.}(2009)\citenamefont
  {Schlegel}, \citenamefont {Naumova}, \citenamefont {Tikhonchuk},
  \citenamefont {Labaune}, \citenamefont {Sokolov},\ and\ \citenamefont
  {Mourou}}]{Schlegel_POP_2009}%
  \BibitemOpen
  \bibfield  {author} {\bibinfo {author} {\bibfnamefont {T.}~\bibnamefont
  {Schlegel}}, \bibinfo {author} {\bibfnamefont {N.}~\bibnamefont {Naumova}},
  \bibinfo {author} {\bibfnamefont {V.~T.}\ \bibnamefont {Tikhonchuk}},
  \bibinfo {author} {\bibfnamefont {C.}~\bibnamefont {Labaune}}, \bibinfo
  {author} {\bibfnamefont {I.~V.}\ \bibnamefont {Sokolov}}, \ and\ \bibinfo
  {author} {\bibfnamefont {G.}~\bibnamefont {Mourou}},\ }\href {\doibase
  10.1063/1.3196845} {\bibfield  {journal} {\bibinfo  {journal} {Phys.
  Plasmas}\ }\textbf {\bibinfo {volume} {16}},\ \bibinfo {eid} {083103}
  (\bibinfo {year} {2009})}\BibitemShut {NoStop}%
\bibitem [{\citenamefont {Yin}\ \emph {et~al.}(2006)\citenamefont {Yin},
  \citenamefont {Albright}, \citenamefont {Hegelich},\ and\ \citenamefont
  {Fern\'andez}}]{Yin_LPB_2006}%
  \BibitemOpen
  \bibfield  {author} {\bibinfo {author} {\bibfnamefont {L.}~\bibnamefont
  {Yin}}, \bibinfo {author} {\bibfnamefont {B.~J.}\ \bibnamefont {Albright}},
  \bibinfo {author} {\bibfnamefont {B.~M.}\ \bibnamefont {Hegelich}}, \ and\
  \bibinfo {author} {\bibfnamefont {J.~C.}\ \bibnamefont {Fern\'andez}},\
  }\href {\doibase 10.1017/S0263034606060459} {\bibfield  {journal} {\bibinfo
  {journal} {Laser Part. Beams}\ }\textbf {\bibinfo {volume} {24}},\ \bibinfo
  {pages} {291} (\bibinfo {year} {2006})}\BibitemShut {NoStop}%
\bibitem [{\citenamefont {Albright}\ \emph {et~al.}(2007)\citenamefont
  {Albright}, \citenamefont {Yin}, \citenamefont {Bowers}, \citenamefont
  {Hegelich}, \citenamefont {Flippo}, \citenamefont {Kwan},\ and\ \citenamefont
  {Fernandez}}]{Albright_POP_2007}%
  \BibitemOpen
  \bibfield  {author} {\bibinfo {author} {\bibfnamefont {B.~J.}\ \bibnamefont
  {Albright}}, \bibinfo {author} {\bibfnamefont {L.}~\bibnamefont {Yin}},
  \bibinfo {author} {\bibfnamefont {K.~J.}\ \bibnamefont {Bowers}}, \bibinfo
  {author} {\bibfnamefont {B.~M.}\ \bibnamefont {Hegelich}}, \bibinfo {author}
  {\bibfnamefont {K.~A.}\ \bibnamefont {Flippo}}, \bibinfo {author}
  {\bibfnamefont {T.~J.~T.}\ \bibnamefont {Kwan}}, \ and\ \bibinfo {author}
  {\bibfnamefont {J.~C.}\ \bibnamefont {Fernandez}},\ }\href {\doibase
  10.1063/1.2768933} {\bibfield  {journal} {\bibinfo  {journal} {Phys.
  Plasmas}\ }\textbf {\bibinfo {volume} {14}},\ \bibinfo {eid} {094502}
  (\bibinfo {year} {2007})}\BibitemShut {NoStop}%
\bibitem [{\citenamefont {Yin}\ \emph {et~al.}(2011)\citenamefont {Yin},
  \citenamefont {Albright}, \citenamefont {Bowers}, \citenamefont {Jung},
  \citenamefont {Fern\'andez},\ and\ \citenamefont {Hegelich}}]{Yin_PRL_2011}%
  \BibitemOpen
  \bibfield  {author} {\bibinfo {author} {\bibfnamefont {L.}~\bibnamefont
  {Yin}}, \bibinfo {author} {\bibfnamefont {B.~J.}\ \bibnamefont {Albright}},
  \bibinfo {author} {\bibfnamefont {K.~J.}\ \bibnamefont {Bowers}}, \bibinfo
  {author} {\bibfnamefont {D.}~\bibnamefont {Jung}}, \bibinfo {author}
  {\bibfnamefont {J.~C.}\ \bibnamefont {Fern\'andez}}, \ and\ \bibinfo {author}
  {\bibfnamefont {B.~M.}\ \bibnamefont {Hegelich}},\ }\href {\doibase
  10.1103/PhysRevLett.107.045003} {\bibfield  {journal} {\bibinfo  {journal}
  {Phys. Rev. Lett.}\ }\textbf {\bibinfo {volume} {107}},\ \bibinfo {pages}
  {045003} (\bibinfo {year} {2011})}\BibitemShut {NoStop}%
\bibitem [{\citenamefont {Max}\ and\ \citenamefont {Perkins}(1971)}]{max1971}%
  \BibitemOpen
  \bibfield  {author} {\bibinfo {author} {\bibfnamefont {C.}~\bibnamefont
  {Max}}\ and\ \bibinfo {author} {\bibfnamefont {F.}~\bibnamefont {Perkins}},\
  }\href {\doibase 10.1103/PhysRevLett.27.1342} {\bibfield  {journal} {\bibinfo
   {journal} {Phys. Rev. Lett.}\ }\textbf {\bibinfo {volume} {27}},\ \bibinfo
  {pages} {1342} (\bibinfo {year} {1971})}\BibitemShut {NoStop}%
\bibitem [{\citenamefont {Marburger}\ and\ \citenamefont
  {Tooper}(1975)}]{marburger1975}%
  \BibitemOpen
  \bibfield  {author} {\bibinfo {author} {\bibfnamefont {J.~H.}\ \bibnamefont
  {Marburger}}\ and\ \bibinfo {author} {\bibfnamefont {R.~F.}\ \bibnamefont
  {Tooper}},\ }\href {\doibase 10.1103/PhysRevLett.35.1001} {\bibfield
  {journal} {\bibinfo  {journal} {Phys. Rev. Lett.}\ }\textbf {\bibinfo
  {volume} {35}},\ \bibinfo {pages} {1001} (\bibinfo {year}
  {1975})}\BibitemShut {NoStop}%
\bibitem [{\citenamefont {Lai}(1976)}]{Lai1976}%
  \BibitemOpen
  \bibfield  {author} {\bibinfo {author} {\bibfnamefont {C.~S.}\ \bibnamefont
  {Lai}},\ }\href {\doibase 10.1103/PhysRevLett.36.966} {\bibfield  {journal}
  {\bibinfo  {journal} {Phys. Rev. Lett.}\ }\textbf {\bibinfo {volume} {36}},\
  \bibinfo {pages} {966} (\bibinfo {year} {1976})}\BibitemShut {NoStop}%
\bibitem [{\citenamefont {Cattani}\ \emph {et~al.}(2000)\citenamefont
  {Cattani}, \citenamefont {Kim}, \citenamefont {Anderson},\ and\ \citenamefont
  {Lisak}}]{cattani2000}%
  \BibitemOpen
  \bibfield  {author} {\bibinfo {author} {\bibfnamefont {F.}~\bibnamefont
  {Cattani}}, \bibinfo {author} {\bibfnamefont {A.}~\bibnamefont {Kim}},
  \bibinfo {author} {\bibfnamefont {D.}~\bibnamefont {Anderson}}, \ and\
  \bibinfo {author} {\bibfnamefont {M.}~\bibnamefont {Lisak}},\ }\href
  {\doibase 10.1103/PhysRevE.62.1234} {\bibfield  {journal} {\bibinfo
  {journal} {Phys. Rev. E}\ }\textbf {\bibinfo {volume} {62}},\ \bibinfo
  {pages} {1234} (\bibinfo {year} {2000})}\BibitemShut {NoStop}%
\bibitem [{\citenamefont {Goloviznin}\ and\ \citenamefont
  {Schep}(2000)}]{goloviznin2000}%
  \BibitemOpen
  \bibfield  {author} {\bibinfo {author} {\bibfnamefont {V.~V.}\ \bibnamefont
  {Goloviznin}}\ and\ \bibinfo {author} {\bibfnamefont {T.~J.}\ \bibnamefont
  {Schep}},\ }\href {\doibase doi:10.1063/1.873976} {\bibfield  {journal}
  {\bibinfo  {journal} {Phys. Plasmas}\ }\textbf {\bibinfo {volume} {7}},\
  \bibinfo {pages} {1564} (\bibinfo {year} {2000})}\BibitemShut {NoStop}%
\bibitem [{\citenamefont {Eremin}\ \emph {et~al.}(2010)\citenamefont {Eremin},
  \citenamefont {Korzhimanov},\ and\ \citenamefont {Kim}}]{eremin2010}%
  \BibitemOpen
  \bibfield  {author} {\bibinfo {author} {\bibfnamefont {V.~I.}\ \bibnamefont
  {Eremin}}, \bibinfo {author} {\bibfnamefont {A.~V.}\ \bibnamefont
  {Korzhimanov}}, \ and\ \bibinfo {author} {\bibfnamefont {A.~V.}\ \bibnamefont
  {Kim}},\ }\href {\doibase doi:10.1063/1.3368791} {\bibfield  {journal}
  {\bibinfo  {journal} {Phys. Plasmas}\ }\textbf {\bibinfo {volume} {17}},\
  \bibinfo {pages} {043102} (\bibinfo {year} {2010})}\BibitemShut {NoStop}%
\bibitem [{\citenamefont {Wilks}\ \emph {et~al.}(1993)\citenamefont {Wilks},
  \citenamefont {Kruer},\ and\ \citenamefont {Mori}}]{wilks1993}%
  \BibitemOpen
  \bibfield  {author} {\bibinfo {author} {\bibfnamefont {S.}~\bibnamefont
  {Wilks}}, \bibinfo {author} {\bibfnamefont {W.}~\bibnamefont {Kruer}}, \ and\
  \bibinfo {author} {\bibfnamefont {W.}~\bibnamefont {Mori}},\ }\href {\doibase
  10.1109/27.221110} {\bibfield  {journal} {\bibinfo  {journal} {{IEEE} T.
  Plasma Sci.}\ }\textbf {\bibinfo {volume} {21}},\ \bibinfo {pages} {120 }
  (\bibinfo {year} {1993})}\BibitemShut {NoStop}%
\bibitem [{\citenamefont {Lefebvre}\ and\ \citenamefont
  {Bonnaud}(1995)}]{lefebvre1995}%
  \BibitemOpen
  \bibfield  {author} {\bibinfo {author} {\bibfnamefont {E.}~\bibnamefont
  {Lefebvre}}\ and\ \bibinfo {author} {\bibfnamefont {G.}~\bibnamefont
  {Bonnaud}},\ }\href {\doibase 10.1103/PhysRevLett.74.2002} {\bibfield
  {journal} {\bibinfo  {journal} {Phys. Rev. Lett.}\ }\textbf {\bibinfo
  {volume} {74}},\ \bibinfo {pages} {2002} (\bibinfo {year}
  {1995})}\BibitemShut {NoStop}%
\bibitem [{\citenamefont {Gu\'erin}\ \emph {et~al.}(1996)\citenamefont
  {Gu\'erin}, \citenamefont {Mora}, \citenamefont {Adam}, \citenamefont
  {H\'eron},\ and\ \citenamefont {Laval}}]{guerin1996}%
  \BibitemOpen
  \bibfield  {author} {\bibinfo {author} {\bibfnamefont {S.}~\bibnamefont
  {Gu\'erin}}, \bibinfo {author} {\bibfnamefont {P.}~\bibnamefont {Mora}},
  \bibinfo {author} {\bibfnamefont {J.~C.}\ \bibnamefont {Adam}}, \bibinfo
  {author} {\bibfnamefont {A.}~\bibnamefont {H\'eron}}, \ and\ \bibinfo
  {author} {\bibfnamefont {G.}~\bibnamefont {Laval}},\ }\href {\doibase
  doi:10.1063/1.871526} {\bibfield  {journal} {\bibinfo  {journal} {Phys.
  Plasmas}\ }\textbf {\bibinfo {volume} {3}},\ \bibinfo {pages} {2693}
  (\bibinfo {year} {1996})}\BibitemShut {NoStop}%
\bibitem [{\citenamefont {Sakagami}\ and\ \citenamefont
  {Mima}(1996)}]{sakagami1996}%
  \BibitemOpen
  \bibfield  {author} {\bibinfo {author} {\bibfnamefont {H.}~\bibnamefont
  {Sakagami}}\ and\ \bibinfo {author} {\bibfnamefont {K.}~\bibnamefont
  {Mima}},\ }\href {\doibase 10.1103/PhysRevE.54.1870} {\bibfield  {journal}
  {\bibinfo  {journal} {Phys. Rev. E}\ }\textbf {\bibinfo {volume} {54}},\
  \bibinfo {pages} {1870} (\bibinfo {year} {1996})}\BibitemShut {NoStop}%
\bibitem [{\citenamefont {Gibbon}(2005)}]{gibbon2005}%
  \BibitemOpen
  \bibfield  {author} {\bibinfo {author} {\bibfnamefont {P.}~\bibnamefont
  {Gibbon}},\ }\href@noop {} {\emph {\bibinfo {title} {Short Pulse Laser
  Interactions with Matter}}}\ (\bibinfo  {publisher} {Imperial College
  Press},\ \bibinfo {address} {London},\ \bibinfo {year} {2005})\BibitemShut
  {NoStop}%
\bibitem [{Note2()}]{Note2}%
  \BibitemOpen
  \bibinfo {note} {Equation~(\ref {eq:n_cel}) then implies $n_e(x)\rightarrow
  n_0$, as $x\rightarrow \infty $.}\BibitemShut {Stop}%
\bibitem [{Note3()}]{Note3}%
  \BibitemOpen
  \bibinfo {note} {The difference $m-k$ corresponds to the number of equilibria
  in the CSL, which is \protect \emph {a priori} unknown.}\BibitemShut {Stop}%
\bibitem [{\citenamefont {Lichtenberg}\ and\ \citenamefont
  {Lieberman}(1992)}]{lichtenberg1992}%
  \BibitemOpen
  \bibfield  {author} {\bibinfo {author} {\bibfnamefont {A.~J.}\ \bibnamefont
  {Lichtenberg}}\ and\ \bibinfo {author} {\bibfnamefont {M.~A.}\ \bibnamefont
  {Lieberman}},\ }\href@noop {} {\emph {\bibinfo {title} {Regular and Chaotic
  Dynamics}}}\ (\bibinfo  {publisher} {Springer},\ \bibinfo {address} {New
  York},\ \bibinfo {year} {1992})\BibitemShut {NoStop}%
\bibitem [{\citenamefont {Birdsall}\ and\ \citenamefont
  {Langdon}(1991)}]{birdsall}%
  \BibitemOpen
  \bibfield  {author} {\bibinfo {author} {\bibfnamefont {C.}~\bibnamefont
  {Birdsall}}\ and\ \bibinfo {author} {\bibfnamefont {A.}~\bibnamefont
  {Langdon}},\ }\href@noop {} {\emph {\bibinfo {title} {Plasma Physics Via
  Computer Simulation}}}\ (\bibinfo  {publisher} {Adam-Hilger},\ \bibinfo
  {address} {Bristol, UK},\ \bibinfo {year} {1991})\BibitemShut {NoStop}%
\bibitem [{\citenamefont {Grech}\ \emph {et~al.}(2012)\citenamefont {Grech},
  \citenamefont {Siminos},\ and\ \citenamefont {Skupin}}]{grech2020}%
  \BibitemOpen
  \bibfield  {author} {\bibinfo {author} {\bibfnamefont {M.}~\bibnamefont
  {Grech}}, \bibinfo {author} {\bibfnamefont {E.}~\bibnamefont {Siminos}}, \
  and\ \bibinfo {author} {\bibfnamefont {S.}~\bibnamefont {Skupin}},\
  }\href@noop {} {} (\bibinfo {year} {2012}),\ \bibinfo {note} {in
  preparation}\BibitemShut {NoStop}%
\bibitem [{\citenamefont {Taflove}\ and\ \citenamefont
  {Hagness}(2005)}]{taflove}%
  \BibitemOpen
  \bibfield  {author} {\bibinfo {author} {\bibfnamefont {A.}~\bibnamefont
  {Taflove}}\ and\ \bibinfo {author} {\bibfnamefont {S.~C.}\ \bibnamefont
  {Hagness}},\ }\href@noop {} {\emph {\bibinfo {title} {Computational
  Electrodynamics: The Finite-Difference Time-Domain Method}}},\ \bibinfo
  {edition} {3rd}\ ed.\ (\bibinfo  {publisher} {Artech House},\ \bibinfo
  {address} {Norwood, MA},\ \bibinfo {year} {2005})\BibitemShut {NoStop}%
\bibitem [{\citenamefont {Boris}(1970)}]{boris1970}%
  \BibitemOpen
  \bibfield  {author} {\bibinfo {author} {\bibfnamefont {J.}~\bibnamefont
  {Boris}},\ }in\ \href@noop {} {\emph {\bibinfo {booktitle} {Proc. Fourth
  Conf. on Numerical Simulation of Plasmas}}}\ (\bibinfo  {publisher} {Naval
  Res. Lab},\ \bibinfo {address} {Washington, D.C.},\ \bibinfo {year} {1970})\
  pp.\ \bibinfo {pages} {3--67}\BibitemShut {NoStop}%
\bibitem [{\citenamefont {Esirkepov}(2001)}]{esirkepov_CPC_2001}%
  \BibitemOpen
  \bibfield  {author} {\bibinfo {author} {\bibfnamefont {T.}~\bibnamefont
  {Esirkepov}},\ }\href {\doibase 10.1016/S0010-4655(00)00228-9} {\bibfield
  {journal} {\bibinfo  {journal} {Comput. Phys. Commun.}\ }\textbf {\bibinfo
  {volume} {135}},\ \bibinfo {pages} {144 } (\bibinfo {year}
  {2001})}\BibitemShut {NoStop}%
\bibitem [{\citenamefont {Sanz}\ \emph {et~al.}(2012)\citenamefont {Sanz},
  \citenamefont {Debayle},\ and\ \citenamefont {Mima}}]{sanz2012}%
  \BibitemOpen
  \bibfield  {author} {\bibinfo {author} {\bibfnamefont {J.}~\bibnamefont
  {Sanz}}, \bibinfo {author} {\bibfnamefont {A.}~\bibnamefont {Debayle}}, \
  and\ \bibinfo {author} {\bibfnamefont {K.}~\bibnamefont {Mima}},\ }\href
  {\doibase 10.1103/PhysRevE.85.046411} {\bibfield  {journal} {\bibinfo
  {journal} {Phys. Rev. E}\ }\textbf {\bibinfo {volume} {85}},\ \bibinfo
  {pages} {046411} (\bibinfo {year} {2012})}\BibitemShut {NoStop}%
\bibitem [{\citenamefont {Giulietti}\ \emph {et~al.}(1997)\citenamefont
  {Giulietti}, \citenamefont {Gizzi}, \citenamefont {Giulietti}, \citenamefont
  {Macchi}, \citenamefont {Teychenn\'e}, \citenamefont {Chessa}, \citenamefont
  {Rousse}, \citenamefont {Cheriaux}, \citenamefont {Chambaret},\ and\
  \citenamefont {Darpentigny}}]{giulietti1997}%
  \BibitemOpen
  \bibfield  {author} {\bibinfo {author} {\bibfnamefont {D.}~\bibnamefont
  {Giulietti}}, \bibinfo {author} {\bibfnamefont {L.~A.}\ \bibnamefont
  {Gizzi}}, \bibinfo {author} {\bibfnamefont {A.}~\bibnamefont {Giulietti}},
  \bibinfo {author} {\bibfnamefont {A.}~\bibnamefont {Macchi}}, \bibinfo
  {author} {\bibfnamefont {D.}~\bibnamefont {Teychenn\'e}}, \bibinfo {author}
  {\bibfnamefont {P.}~\bibnamefont {Chessa}}, \bibinfo {author} {\bibfnamefont
  {A.}~\bibnamefont {Rousse}}, \bibinfo {author} {\bibfnamefont
  {G.}~\bibnamefont {Cheriaux}}, \bibinfo {author} {\bibfnamefont {J.~P.}\
  \bibnamefont {Chambaret}}, \ and\ \bibinfo {author} {\bibfnamefont
  {G.}~\bibnamefont {Darpentigny}},\ }\href {\doibase
  10.1103/PhysRevLett.79.3194} {\bibfield  {journal} {\bibinfo  {journal}
  {Phys. Rev. Lett.}\ }\textbf {\bibinfo {volume} {79}},\ \bibinfo {pages}
  {3194} (\bibinfo {year} {1997})}\BibitemShut {NoStop}%
\bibitem [{\citenamefont {Fuchs}\ \emph {et~al.}(1998)\citenamefont {Fuchs},
  \citenamefont {Adam}, \citenamefont {Amiranoff}, \citenamefont {Baton},
  \citenamefont {Gallant}, \citenamefont {Gremillet}, \citenamefont {H\'eron},
  \citenamefont {Kieffer}, \citenamefont {Laval}, \citenamefont {Malka},
  \citenamefont {Miquel}, \citenamefont {Mora}, \citenamefont {P\'epin},\ and\
  \citenamefont {Rousseaux}}]{fuchs1998}%
  \BibitemOpen
  \bibfield  {author} {\bibinfo {author} {\bibfnamefont {J.}~\bibnamefont
  {Fuchs}}, \bibinfo {author} {\bibfnamefont {J.~C.}\ \bibnamefont {Adam}},
  \bibinfo {author} {\bibfnamefont {F.}~\bibnamefont {Amiranoff}}, \bibinfo
  {author} {\bibfnamefont {S.~D.}\ \bibnamefont {Baton}}, \bibinfo {author}
  {\bibfnamefont {P.}~\bibnamefont {Gallant}}, \bibinfo {author} {\bibfnamefont
  {L.}~\bibnamefont {Gremillet}}, \bibinfo {author} {\bibfnamefont
  {A.}~\bibnamefont {H\'eron}}, \bibinfo {author} {\bibfnamefont {J.~C.}\
  \bibnamefont {Kieffer}}, \bibinfo {author} {\bibfnamefont {G.}~\bibnamefont
  {Laval}}, \bibinfo {author} {\bibfnamefont {G.}~\bibnamefont {Malka}},
  \bibinfo {author} {\bibfnamefont {J.~L.}\ \bibnamefont {Miquel}}, \bibinfo
  {author} {\bibfnamefont {P.}~\bibnamefont {Mora}}, \bibinfo {author}
  {\bibfnamefont {H.}~\bibnamefont {P\'epin}}, \ and\ \bibinfo {author}
  {\bibfnamefont {C.}~\bibnamefont {Rousseaux}},\ }\href {\doibase
  10.1103/PhysRevLett.80.2326} {\bibfield  {journal} {\bibinfo  {journal}
  {Phys. Rev. Lett.}\ }\textbf {\bibinfo {volume} {80}},\ \bibinfo {pages}
  {2326} (\bibinfo {year} {1998})}\BibitemShut {NoStop}%
\bibitem [{\citenamefont {Willingale}\ \emph {et~al.}(2009)\citenamefont
  {Willingale}, \citenamefont {Nagel}, \citenamefont {Thomas}, \citenamefont
  {Bellei}, \citenamefont {Clarke}, \citenamefont {Dangor}, \citenamefont
  {Heathcote}, \citenamefont {Kaluza}, \citenamefont {Kamperidis},
  \citenamefont {Kneip}, \citenamefont {Krushelnick}, \citenamefont {Lopes},
  \citenamefont {Mangles}, \citenamefont {Nazarov}, \citenamefont {Nilson},\
  and\ \citenamefont {Najmudin}}]{willingale2009}%
  \BibitemOpen
  \bibfield  {author} {\bibinfo {author} {\bibfnamefont {L.}~\bibnamefont
  {Willingale}}, \bibinfo {author} {\bibfnamefont {S.~R.}\ \bibnamefont
  {Nagel}}, \bibinfo {author} {\bibfnamefont {A.~G.~R.}\ \bibnamefont
  {Thomas}}, \bibinfo {author} {\bibfnamefont {C.}~\bibnamefont {Bellei}},
  \bibinfo {author} {\bibfnamefont {R.~J.}\ \bibnamefont {Clarke}}, \bibinfo
  {author} {\bibfnamefont {A.~E.}\ \bibnamefont {Dangor}}, \bibinfo {author}
  {\bibfnamefont {R.}~\bibnamefont {Heathcote}}, \bibinfo {author}
  {\bibfnamefont {M.~C.}\ \bibnamefont {Kaluza}}, \bibinfo {author}
  {\bibfnamefont {C.}~\bibnamefont {Kamperidis}}, \bibinfo {author}
  {\bibfnamefont {S.}~\bibnamefont {Kneip}}, \bibinfo {author} {\bibfnamefont
  {K.}~\bibnamefont {Krushelnick}}, \bibinfo {author} {\bibfnamefont
  {N.}~\bibnamefont {Lopes}}, \bibinfo {author} {\bibfnamefont {S.~P.~D.}\
  \bibnamefont {Mangles}}, \bibinfo {author} {\bibfnamefont {W.}~\bibnamefont
  {Nazarov}}, \bibinfo {author} {\bibfnamefont {P.~M.}\ \bibnamefont {Nilson}},
  \ and\ \bibinfo {author} {\bibfnamefont {Z.}~\bibnamefont {Najmudin}},\
  }\href {\doibase 10.1103/PhysRevLett.102.125002} {\bibfield  {journal}
  {\bibinfo  {journal} {Phys. Rev. Lett.}\ }\textbf {\bibinfo {volume} {102}},\
  \bibinfo {pages} {125002} (\bibinfo {year} {2009})}\BibitemShut {NoStop}%
\bibitem [{\citenamefont {Palaniyappan}\ \emph {et~al.}(2012)\citenamefont
  {Palaniyappan}, \citenamefont {Hegelich}, \citenamefont {Wu}, \citenamefont
  {Jung}, \citenamefont {Gautier}, \citenamefont {Yin}, \citenamefont
  {Albright}, \citenamefont {Johnson}, \citenamefont {Shimada}, \citenamefont
  {Letzring}, \citenamefont {Offermann}, \citenamefont {Ren}, \citenamefont
  {Huang}, \citenamefont {H\"orlein}, \citenamefont {Dromey}, \citenamefont
  {Fernandez},\ and\ \citenamefont {Shah}}]{palaniyappan2012}%
  \BibitemOpen
  \bibfield  {author} {\bibinfo {author} {\bibfnamefont {S.}~\bibnamefont
  {Palaniyappan}}, \bibinfo {author} {\bibfnamefont {B.~M.}\ \bibnamefont
  {Hegelich}}, \bibinfo {author} {\bibfnamefont {H.-C.}\ \bibnamefont {Wu}},
  \bibinfo {author} {\bibfnamefont {D.}~\bibnamefont {Jung}}, \bibinfo {author}
  {\bibfnamefont {D.~C.}\ \bibnamefont {Gautier}}, \bibinfo {author}
  {\bibfnamefont {L.}~\bibnamefont {Yin}}, \bibinfo {author} {\bibfnamefont
  {B.~J.}\ \bibnamefont {Albright}}, \bibinfo {author} {\bibfnamefont {R.~P.}\
  \bibnamefont {Johnson}}, \bibinfo {author} {\bibfnamefont {T.}~\bibnamefont
  {Shimada}}, \bibinfo {author} {\bibfnamefont {S.}~\bibnamefont {Letzring}},
  \bibinfo {author} {\bibfnamefont {D.~T.}\ \bibnamefont {Offermann}}, \bibinfo
  {author} {\bibfnamefont {J.}~\bibnamefont {Ren}}, \bibinfo {author}
  {\bibfnamefont {C.}~\bibnamefont {Huang}}, \bibinfo {author} {\bibfnamefont
  {R.}~\bibnamefont {H\"orlein}}, \bibinfo {author} {\bibfnamefont
  {B.}~\bibnamefont {Dromey}}, \bibinfo {author} {\bibfnamefont {J.~C.}\
  \bibnamefont {Fernandez}}, \ and\ \bibinfo {author} {\bibfnamefont {R.~C.}\
  \bibnamefont {Shah}},\ }\href {\doibase 10.1038/nphys2390} {\bibfield
  {journal} {\bibinfo  {journal} {Nature Phys.}\ }\textbf {\bibinfo {volume}
  {8}},\ \bibinfo {pages} {763} (\bibinfo {year} {2012})}\BibitemShut {NoStop}%
\end{thebibliography}
\end{document}